\documentclass[12pt,reqno]{article}
\usepackage{chet}
\usepackage{booktabs}
\usepackage{amsmath,hyperref,amssymb}
\pdfoutput=1
\usepackage{graphicx,url}
\usepackage{color}
\usepackage{setspace}
 
\usepackage{epsfig}

\def\tr{{\rm Tr}}

\def\Or[#1]{{\text{O}}\left({#1}\right)}
\def\dotl[#1,#2]{\left\langle #1, #2 \right\rangle}
\def\dotlb[#1,#2]{[ #1, #2 ]}
\def\dotp[#1,#2]{(#1) \cdot (#2)}
\def\aff[#1,#2]{\hat{#1}(#2)}
\def\n4sym{{\cal N}=4 SYM}
\def\>{\rangle}
\def\<{\langle}
\def\weight[#1,#2,#3]{\{(#1),#2,#3\}}
\def\ads[#1]{$\text{AdS}_{#1}$}

\newcommand{\ba}{\begin{eqnarray}}
\newcommand{\ea}{\end{eqnarray}}
\newcommand{\be}{\begin{eqnarray}}
\newcommand{\ee}{\end{eqnarray}}
\newcommand{\bq}{\begin{equation}}
\newcommand{\eq}{\end{equation}}
\newcommand{\benn}{\begin{equation*}}
\newcommand{\eenn}{\end{equation*}}
\newcommand{\bi}{\begin{itemize}}  
\newcommand{\ei}{\end{itemize}}

\newcommand{\CL}{{\cal L}}

\newcommand{\CM}{{\cal M}}

\newcommand{\CO}{{\cal O}}
\newcommand{\CP}{{\cal P}}

\newcommand{\CV}{{\cal V}}
\newcommand{\nn}{\nonumber}

\newcommand\oo\infty
\newcommand\s\sigma
\newcommand\de\delta
\newcommand\De\Delta

\newcommand\f\phi
\newcommand\g\gamma
\newcommand\x\times

\begin{document}

\title{Conformal Blocks Beyond the Semi-Classical Limit}%/ Ondes Partielles Conformes Au-del\`a La Limite Semi-Classique}
\author{A. Liam Fitzpatrick$^1$, Jared Kaplan$^2$}
\affiliation{
{\it $^1$ Dept.\ of Physics, Boston University, Boston, MA 02215} \\
{\it $^2$ Dept. of Physics and Astronomy, Johns Hopkins University, Baltimore, MD 21218} \\
}
\abstract{ 

Black hole microstates and their approximate thermodynamic properties can be studied using heavy-light correlation functions in AdS/CFT.  Universal features of these correlators can be extracted from the Virasoro conformal blocks in CFT$_2$, which encapsulate quantum gravitational effects in AdS$_3$.  At infinite central charge $c$, the Virasoro vacuum block provides an avatar of the black hole information paradox in the form of periodic Euclidean-time singularities that must be resolved at finite $c$.  

We compute Virasoro blocks in the heavy-light, large $c$ limit, extending our previous results by determining perturbative $1/c$ corrections.  We obtain explicit closed-form expressions for both the `semi-classical' $h_L^2 / c^2$ and `quantum' $h_L / c^2$ corrections to the vacuum block, and we provide integral formulas for general Virasoro blocks.  We comment on the interpretation of our results for thermodynamics, discussing how monodromies in Euclidean time can arise from AdS calculations using `geodesic Witten diagrams'.  We expect that only non-perturbative corrections in $1/c$ can resolve the singularities associated with the information paradox.
}

\maketitle

\tableofcontents

\flushbottom
 
\section{Introduction and Discussion}
 
To make predictions about the thermodynamic behavior of a system, we usually study a statistical ensemble of states codified by a partition function.  In this standard `macroscopic' approach, the entropy function $S(E)$ plays a key role, counting the number of states $e^{S(E)}$ with energy $E$ and determining the phase diagram of the theory as a function of the temperature.  For example, the Cardy formula \cite{Cardy:1986ie, Hartman:2014oaa} for $S(E)$ predicts the asymptotic density of states in CFT$_2$, thereby counting the number of black hole states in quantum gravity theories in AdS$_3$ \cite{Strominger:1996sh, Strominger:1997eq}. 

We have taken a rather different `microscopic' approach to thermodynamics in AdS/CFT \cite{Maldacena:1997re,Witten, GKP},  studying the correlation functions of light probe operators in the background of a heavy CFT microstate.  Intuitively, we expect that there should be very little difference between observables computed in a thermal density matrix and those computed in a pure state randomly chosen from the canonical  ensemble.  Via the operator/state correspondence, we can infer thermodynamic properties from a 4-pt correlator by comparing
\be \label{eq:ThermalEqualsPure1}
 \< \CO_H(\infty)  \CO_L(1) \CO_L(z)  \CO_H(0) \> \overset{?}{\approx} \< \CO_L(1) \CO_L(z) \>_{T_H} = \left( \frac{\pi T_H}{\sin(\pi T_H t)} \right)^{2h_L}
\ee
where $z = 1- e^{-t}$, $\CO_H$ is a heavy operator, and the last equality holds in CFT$_2$.  We obtain precisely this relation \cite{Fitzpatrick:2014vua} by approximating the left-hand side with the Virasoro vacuum conformal block, computed at large central charge $c$ in the limit $h_H \propto c \gg h_L$.  In the light-cone OPE limit \cite{Fitzpatrick:2012yx, KomargodskiZhiboedov}, this will be a good approximation for any CFT$_2$ without additional conserved currents; more generally it provides an interesting universal contribution to the correlator capturing gravitational effects in AdS$_3$.  Thus the thermodynamic properties of high energy states in CFT$_2$ at large $c$ are built into the structure of the Virasoro algebra.  

In this work we will study $1/c$ corrections to the Virasoro conformal blocks and their implications for thermodynamics.  These will include both semi-classical corrections at higher orders in $h_L / c$ and genuine `quantum' corrections.  We use the terminology `semi-classical' and `quantum' because these correspond, respectively, to the gravitational backreaction of the light probe and to gravitational loop effects in AdS$_3$.   In the remainder of this introduction we will discuss how our discussion relates to the black hole information paradox, and then we provide a summary of the results.  

Chaos can also be studied by taking a limit of CFT 4-point correlators \cite{Shenker:2013pqa, Roberts:2014ifa, Shenker:2014cwa}, with a universal bound expected for large central charge theories \cite{Maldacena:2015waa}; the implications of our results for chaos will be discussed in a forthcoming work.

\subsection{The Information Paradox and the Vacuum Block}

The black hole information paradox has many guises.  In its most visceral and pressing form, it requires understanding the correct description of physics near black hole horizons, and in particular, the question of whether the semi-classical description can survive as a good approximation while simultaneously allowing for unitary evolution \cite{Mathur:2009hf,Mathur:2010kx, Almheiri:2012rt,Braunstein:2009my}.  Such problems remain extremely perplexing and important, but they are difficult (or impossible?) to formulate as a precise question about CFT observables, and progress on this front may  require qualitatively new `observables' \cite{Papadodimas:2012aq, Papadodimas:2013jku, Harlow:2014yoa}.

A more straightforward manifestation of the information paradox can be formulated directly in terms of CFT correlators.  In the background of a large AdS-Schwarzschild black hole, the two-point correlation function with Lorentzian time-separation $t_L$ decays exponentially at large time \cite{Maldacena:2001kr}.  This means that information dropped into the black hole at an initial time never comes out.   A CFT living on a non-compact space or a CFT at infinite central charge may also have thermal correlators that decay exponentially for all times, as can be seen explicitly by analytically continuing the right-hand side of equation (\ref{eq:ThermalEqualsPure1}) for the case of CFT$_2$ on the thermal cylinder.  
However, for a CFT living on a compact space with finite central charge and at a finite temperature,\footnote{For a CFT, we can connect the non-compact and compact cases by taking the infinite temperature limit and measuring distances in units of $1/T$.} correlators cannot decay exponentially for all times, as this would signal loss of information concerning a perturbation to the thermal density matrix.

We add another layer to the story by studying the correlators of light operators in the background of a heavy pure state.  This makes it possible to probe the pure quantum state of a one-sided BTZ black hole, instead of an ensemble of $e^S$ black holes.  In the thermodynamic limit we expect the relation of equation (\ref{eq:ThermalEqualsPure1}) to hold, leading to a sharp Euclidean-time signature of information loss.
Thermal 2-pt correlators are periodic under $t_E \to t_E + \beta$.  This periodicity leads to additional singularities in  
equation (\ref{eq:ThermalEqualsPure1}) from periodic images of the $\CO_L(z) \CO_L(1)$ OPE singularity, which occur in the Euclidean region at $z = \bar z = 1 - e^{\frac{n}{T_H}}$ for any integer $n$.  Although these singularities are obligatory for thermal 2-pt correlators, they are \emph{forbidden} in the 4-pt correlators of a CFT at finite central charge $c$ \cite{OPEConvergence}.  So these singularities are a sharp signature of information loss in the large central charge limit, analogous to the bulk point singularity \cite{GGP, JP, Maldacena:2015iua}, a signature of bulk locality.   

In the case of either exponential decay in $t_L$ for thermal 2-pt correlators or periodicity in $t_E$ for pure-state 4-pt correlators, it would be most interesting to have a bulk computation resolving the paradox.  Unfortunately, we do not have a non-perturbative definition of the bulk theory, and in fact, the bulk theory may be precisely defined only via a dual CFT.  

In this paper we will focus on Euclidean time periodicity and its manifestation in the Virasoro conformal blocks.  We expect that unitarity can only be restored by non-perturbative effects in $1/c$, and in particular that perturbative $1/c$ corrections should not violate the thermal periodicity $t_E \to t_E + \beta$ of the large $c$ heavy-light correlators.  These expectations are primarily based on the expectation that $1/c$ corrections correspond to loop effects around the infinite $c$ gravity saddle, which is an AdS black hole background with fixed Euclidean-time periodicity, and thus such corrections should at most produce perturbative corrections to $\beta$.  Roughly speaking, unitarity restoration should rely on contributions from different saddles and therefore involve effects of order $e^{-S} \sim e^{-\CO(c)}$.\footnote{Recall that $S(E) \approx 2 \pi \sqrt{\frac{c E}{6}}$, so for $E \propto c$ this is formally $\CO(c)$. }  Such non-perturbative effects will be addressed more directly in future work.

We will compute $1/c$ corrections to the Virasoro blocks and study their behavior in Euclidean time.  We find that the $1/c$ corrections to the vacuum block do violate periodicity, with a non-trivial monodromy under $t_E \to t_E + \beta$.  Intriguingly, there appear to be two relevant time scales, of order $t \sim c$ and $t \sim e^{\CO(c)}$, as the correlator has non-trivial dependence on both $t$ and $\log t$.  

However, we do not believe that these effects have any immediate connection with the resolution of the information paradox.  Conformal blocks have unphysical monodromies in the Euclidean plane that cancel when they are summed to form full CFT correlation functions.   The monodromies we find in the $1/c$ expansion of Virasoro blocks seem to play a similar role to the more banal monodromies of global conformal blocks.  In section \ref{sec:ExpectationsThermoAdSCFT} we explain how these monodromies can arise from AdS computations of the blocks in terms of `geodesic Witten diagrams' \cite{Hijano:2015zsa, Hijano:2015qja, Hijano:2015rla}.  The case of both global and Virasoro blocks can be given a parallel treatment, which suggests that the Euclidean-time monodromies of the $1/c$ corrections are likely to disappear in the full correlators. 

\subsection{Summary of Results}

In previous work we showed that in the heavy-light semi-classical limit, the vacuum conformal block can be written as
\be
\mathcal{V}(t) = e^{h_L t}  \left( \frac{\pi T_H}{\sin(\pi T_H t)} \right)^{2h_L},
 \ee
where 
\be
 T_H  = \frac{\sqrt{\frac{24 h_H}{c} - 1}}{2 \pi}
\ee
and $z = 1 - e^{-t}$.
Rescaling $\tau \equiv i \pi T_H t$ so that we measure distances in units of $T_H$, and then taking $T_H \to \infty$, we see that the full structure of the vacuum block is preserved. 

Here we show that in the large temperature limit, the first correction in a $1/c$ expansion of the heavy-light vacuum block is 
\begin{equation}
\begin{aligned}
\mathcal{V}(\tau) &= e^{h_L t}  \left( \frac{\pi T_H}{\sin(\pi T_H t)} \right)^{2h_L} \left[ 1 + \frac{h_L}{c} {\cal V}_{h_L}^{(1)} + \frac{h_L^2}{c} {\cal V}_{h_L^2}^{(1)} \right], \\ 
 {\cal V}_{h_L}^{(1)} &= \frac{\text{csch}^2\left(\frac{\alpha  t}{2}\right)}{2} \Big[3 \left(e^{-\alpha t} B\left(e^{-t},-\alpha ,0 \right)+e^{\alpha  t} B\left(e^{-t},\alpha, 0 \right)+e^{\alpha  t}
   B\left(e^t,-\alpha,0 \right)+e^{-\alpha  t} B\left(e^t,\alpha , 0\right)\right)  \\
    &  +\frac{1}{\alpha ^2}+ \cosh (\alpha  t)\left( -\frac{1}{\alpha ^2}+6 H_{-\alpha }+6 H_{\alpha }+6 i \pi -5 \right)+12 \log \left(2 \sinh \left(\frac{t}{2}\right)\right)+ 5\Big] \\
     & -t \frac{\left(13
   \alpha ^2-1\right)  \coth \left(\frac{\alpha  t}{2}\right)}{2\alpha }+12 \log \left(\frac{2 \sinh \left(\frac{\alpha 
   t}{2}\right)}{\alpha }\right),     \\
 {\cal V}_{h_L^2}^{(1)} &=  
 6 \Big( \text{csch}^2\left(  \frac{\alpha  t}{2}\right) \left[ \frac{B(e^{-t},-\alpha,0) +B(e^t,-\alpha ,0)+B(e^{-t},\alpha ,0)+B(e^t,\alpha,0)}{2} \right.  \\
    & \left. + H_{-\alpha }+H_{\alpha }+2 \log \left(2 \sinh \left(\frac{t}{2}\right)\right)+i \pi \right] 
    +2 \left(\log \left(\alpha  \sinh \left(\frac{t}{2}\right) \text{csch}\left(\frac{\alpha  t}{2}\right)\right)+1\right)\Big). 
    \label{eq:ExactResult}
    \end{aligned}
\end{equation}
where  
$B(x,\beta,0) = \frac{x^{\beta}{}_2F_1(1,\beta,1+\beta,x)}{\beta}$ is the incomplete Beta function, 
$z = 1-e^{-t}$,  $H_n$ is the harmonic function, and $\alpha \equiv \sqrt{1- \frac{24 h_H}{c}} \cong 2 \pi i T_H$.\footnote{These expressions have various branch cuts; to be precise, one should start with the conventional definition of these special functions in the region $ \textrm{Im}(z)<0$ to obtain the ``first sheet'' behavior near the Euclidean OPE limit, and extend the function by analytic continuation.}

An important point is that the methods we use in this paper can obtain terms  that are not visible at any order in the ``semi-classical'' part of the conformal blocks.  This semi-classical part is defined as
\be
\lim_{c \rightarrow \infty} \frac{1}{c} \log {\cal V}(z),
\ee
where the ratios $\delta_i \equiv h_i/c$ of the external dimensions to $c$ are all held fixed. After taking the logarithm of ${\cal V}$, the $\CO(h_L^2/c)$ correction term above can be seen to survive in this limit, but the $\CO(h_L/c)$ term does not and thus goes not only beyond leading order in $\delta_L$ but beyond the semi-classical limit itself.

{\em After this work was substantially completed, the paper \cite{Beccaria} appeared that uses a different method to compute an integral expression for the order $h_L^2 / c$ (semi-classical) result. 
}

\section{Corrections to the Vacuum Conformal Block}
\label{sec:CorrectionstoBlock}

\subsection{Review}

Conformal blocks in 2d CFTs are contributions to four-point correlation functions from irreducible representations of the full Virasoro algebra, and as such resum contributions from all powers of the stress tensor.  These contributions are dual to those of all multi-graviton contributions in AdS, and thus automatically encode an enormous amount of information about gravity in AdS$_3$.  To distinguish these conformal blocks from simpler expression that contain irreducible representations of the global subgroup $SL(2, \mathbf{C})$, we refer to the former as Virasoro conformal blocks and the latter as global conformal blocks.  The explicit form for global conformal blocks in 2d has been known for some time and is just a hypergeometric function \cite{ZamolodchikovRecursion}; this is in contrast with Virasoro blocks, where, despite various systematic expansions \cite{ZamolodchikovRecursion,Zamolodchikovq,HartmanLargeC,HarlowLiouville,Alday:2009aq,Perlmutter:2015iya,Headrick}, no closed form expression is known.  In \cite{Fitzpatrick:2014vua,Fitzpatrick:2015zha,Fitzpatrick:2015foa,KrausBlocks,Beccaria,Alkalaev:2015wia,Hijano:2015qja, Alkalaev:2015lca} methods have been developed for computing the Virasoro conformal blocks in a ``heavy-light'' limit, where the central charge as well as the conformal weight of two ``heavy'' external operators are taken to be large, while the conformal weight of two ``light'' external operators is held fixed.  The most efficient technique \cite{Fitzpatrick:2015zha} works by using the conformal anomaly to absorb the leading order contribution of the stress tensor in this limit into a deformation of the metric.

To be more precise, recall that the Laurent coefficients of the stress tensor depend on the coordinates being used:
\be
T(x) &=& \sum_{n=-\infty}^\infty \frac{L^{(x)}_n}{x^{n+2}}.
\ee
The usual Virasoro generators $L_n \equiv L_n^{(z)}$ are the Laurent coefficients in the flat coordinate $z$, where the CFT lives in the metric $ds^2 = dz d \bar z$.  The subset of $L_{n}$ with $n\le -1$ are raising operators which, when acting on a primary state, provide a natural basis for all states in a conformal block.  So, one can work out the conformal blocks for a four-point function $\< \phi_H(\infty) \phi_H(1) \phi_L(z) \phi_L(0)\>$ by expanding the state created by $\phi_L(z) \phi_L(0) | 0 \>$ in this natural basis
\be
\phi_L(z) \phi_L(0) | 0 \> &\supset & z^{h-2h_L}  \sum_{\{ m_i, k_i \}} c_{\{ m_i, k_i  \}} z^{ \sum_i m_i k_i } L_{-m_1}^{k_1} \dots L_{-m_n}^{k_n} |h \>,
\label{eq:BlockOPE}
\ee
where $|h\>$ is the primary state of the conformal block and $c_{\{m_i, k_i \}}$ are coefficients that are fixed by conformal symmetry.  Recall that primary states are defined as those annihilated by the lowering operators $L_{n}$ with $n \geq 1$. One way to compute the Virasoro block is to construct a projector ${\cal P}_h$:
\be
&& {\cal P}_h \equiv \sum_{\{m_i, k_i\}, \{m'_i, k'_i\}}   L_{-m_1}^{k_1} \dots L_{-m_n}^{k_n} | h\> {\cal N}^{-1}_{ \{ m_i, k_i \}, \{m'_i, k'_i \} }  \< h | L_{m'_s}^{k'_s} \dots L_{m'_1}^{k'_1}  , \nn\\
&&{\cal N}_{ \{ m'_i, k'_i \}, \{m_i, k_i \} }\equiv \< h | L_{m'_s}^{k'_s} \dots L_{m'_1}^{k'_1} | L_{-m_1}^{k_1} \dots L_{-m_n}^{k_n}|h \> .
\ee
Acting with ${\cal P}_h$ to make ${\cal P}_h \phi_{L}(z) \phi_{L}(0)| 0\>$, one automatically obtains the sum over the basis in (\ref{eq:BlockOPE}) with coefficients given by
evaluating $\sum_{\{ m'_i, k'_i\}} {\cal N}^{-1}_{ \{ m_i, k_i \}, \{m'_i, k'_i \} } \< h | L_{m'_s}^{k'_s} \dots L_{m'_1}^{k'_1}  \phi_{L}(z) \phi_{L}(0) | 0\>$.  The conformal block itself is just given by $\< \phi_H(\infty) \phi_H(1) {\cal P}_h \phi_L(z) \phi_L(0)\>$, the four point correlator projectioned onto the irreducible representation of Virasoro built from the primary state $| h \>$.  

However, in the heavy-light limit, this is not a very efficient basis to use. Although the normalization factors ${\cal N}_{ \{ m'_i, k'_i \}, \{m_i, k_i \} }$ grow with $c$ for most contributions and thus produce a large suppression, these can be compensated  in the Virasoro block by factors of the heavy operator dimension coming from the numerator  $\< \phi_H(\infty) \phi_H(1)   L_{-m_1}^{k_1} \dots L_{-m_n}^{k_n} | h\>$.  Fortunately,  there exists another natural basis that avoids this difficulty.  It is easy to see that any other set of coordinates $x$ which begins linearly in Euclidean coordinates $z$ at small $z$ will again have the property that $L_n^{(x)}$ with $n\le -1$, and thus also provides a natural basis. In \cite{Fitzpatrick:2015zha} it was noted that the choice of coordinates 
\be
w = 1-(1-z)^\alpha, \qquad \alpha = \sqrt{1 - \frac{24 h_H}{c}},
\ee
leads to remarkable simplifications in the basis generated by ${\cal L}_{-n} \equiv L^{(w)}_{-n}$; in particular, at leading order in $1/c$, the only basis elements that contribute are those of the form ${\cal L}_{-1}^n |h \>$.  The reason is that when one forms the projector ${\cal P}_{h,w}$ in this basis, there is no longer any enhancement from the conformal weight of the heavy operator in $\< \phi_H(\infty) \phi_H(1)   {\cal L}_{-m_1}^{k_1} \dots {\cal L}_{-m_n}^{k_n} | h\>$.  The simplest way to see this is to note that due to the conformal anomaly, 
\be
\frac{\< \phi_H(\infty) \phi_H(1) T(w)| h \>}{\< \phi_H(\infty) \phi_H(1) |h\>} = h \frac{1-z(w)}{z^2(w)} .
\ee
  This does {\em not} grow with $h_H$, and therefore factor of $h_H$ cannot compensate for the suppression by factors of $c$ from the norms ${\cal N}_{ \{ m'_i, k'_i \}, \{m_i, k_i \} }$.

To go to subleading orders in $1/c$, we have to include some of these suppressed terms.   Clearly, we have to include terms where the suppression from the norm involves only one factor of $c$, but there are also some contributions that must be included where the norm produces two factors of $c$.  The reason is that in the sum over modes, factors of the form $\< \phi_H(\infty) \phi_H(1) {\cal L}_{-n} {\cal L}_{-m} | h \>$ with two ${\cal L}$'s can produce a factor of $c$ upstairs.  This is again easiest to understand by looking at correlators with the stress tensor, where this positive factor of $c$ arises from the limit when two $T$'s are brought together.
 In general, a correlator with $2k$ insertions of $T$ can have at most $c^k$ upstairs from such $TT$ OPE singularities, and there will be a suppression by $c^{-2k}$ coming from the norm of the physical $T$ modes.  Thus, to compute to order $1/c^k$ we will have to consider $2k$ factors of ${\cal L}_{-n}$'s.  

\subsection{Computation}
\label{sec:computation}

The projector ${\cal P}_{h,w}$ for the ${\cal L}_{-n}$ is similar to the original Euclidean basis projector ${\cal P}_h$.  Inside a four-point function, it takes the form
\be
&&\< \phi_{H_1}(\infty) \phi_{H_2}(1) {\cal P}_{h,w} \phi_{L_1}(z) \phi_{L_2}(0) \> = \\
 && \sum_{\{m_i, k_i\}, \{m'_i, k'_i\}}  \< \phi_{H_1}(\infty) \phi_{H_2}(1) \CL_{-m_1}^{k_1} \dots \CL_{-m_n}^{k_n} | h\> {\cal N}^{-1}_{ \{ m_i, k_i \}, \{m'_i, k'_i \} }  \< h | \CL_{m'_s}^{k'_s} \dots \CL_{m'_1}^{k'_1} \phi_{L_1}(w) \phi_{L_2}(0)\> .\nn
\ee
As shown in \cite{Fitzpatrick:2015zha}, this correctly acts as a projector onto the ${\cal L}_{-n}$ modes when the overlap factor $\< h | \CL_{m'_s}^{k'_s} \dots \CL_{m'_1}^{k'_1} \phi_L(w) \phi_L(0)\> $ is just given by the analogous Euclidean overlap factor after a conformal transformation on the $\phi_L$'s:
\be
\< h | \CL_{m'_s}^{k'_s} \dots \CL_{m'_1}^{k'_1} \phi_{L_1}(w) \phi_{L_2}(0)\> \equiv (w'(z))^{h_{L_1}} (w'(0))^{h_{L_2}} \< h | L_{m'_s}^{k'_s} \dots L_{m'_1}^{k'_1} \phi_{L_1}(z(w)) \phi_{L_2}(0)\> .
\label{eq:wtrans}
\ee
The norm factors ${\cal N}_{ \{ m_i, k_i \}, \{m'_i, k'_i \} } $ are unchanged from the Euclidean basis. The only piece that changes substantially is the overlap with the heavy operators:
\be
 \< \phi_H(\infty) \phi_H(1) \CL_{-m_1}^{k_1} \dots \CL_{-m_n}^{k_n} | h\> .
 \ee
Our strategy for computing these will be to compute the corresponding $\< \phi_H(\infty) \phi_H(1) T(w_1) \dots T(w_n)\>$ correlators and read off the Laurent coefficients.  In the following, we will focus on the vacuum block with $h_{H_1} = h_{H_2}, h_{L_1}=h_{L_2}$ for simplicity, and relegate the calculation of the general case to appendix \ref{app:gencase}. 

 It will be convenient to choose the insertions of the heavy operators to be at 0 and $\infty$ rather than at 1 and $\infty$; this corresponds to $z\rightarrow 1-z$ and $ w\rightarrow 1-w$ compared to above.  Correlators can be computed in $w$ most easily by using the OPE: 
\be
T(w) \phi_{H}(\infty) &\stackrel{w \sim \infty}{\sim}&  0, \\
T(w) \phi_{H}(0) &\stackrel{w \sim 0}{\sim}&   0 , \nn\\
T(w_1) T(w_2) &\stackrel{w_1 \sim w_2}{\sim}& \frac{c}{\alpha^4 w_1^2 w_2^2} \left( \frac{ z_1^2 z_2^2 }{2 z_{12}^4} + \frac{z_1 z_2}{z_{12}^2} \left( \frac{1-\alpha^2}{12} +\frac{\alpha^2 w_1 w_2 (T(w_1) + T(w_2)) }{c}\right)  \right)  , \nn
\ee
where $h_H = \frac{c}{24} (1 - \alpha^2)$ and $z_i \equiv z(w_i)= w_i^{\frac{1}{\alpha}}$. The notation ``$\sim$'' here means ``equal up to regular terms.'' Since $w^2T(w)$ is holomorphic in $z(w)$, these OPEs determine the singularities and therefore the complete functional dependence of $T$ correlators in terms of correlators without $T$ insertions.  Since the transformation from $w$ to $z$ is regular except at $z=0,\infty$, the last OPE above, $T(w_1) T(w_2)$ is just a rewriting of the standard $TT$ OPE $T(w_1) T(w_2) \sim \frac{c/2}{w_{12}^4} + \frac{2 T(w_2)}{w_{12}^2} + \frac{ \partial_{w_2} T(w_2)}{w_{12}} + \cdots$.  

 Applying the OPE to one or two insertions of $T(w)$ we find
\be
\frac{\< \phi_H(\infty) \phi_H(0) T(w) \>}{\< \phi(\infty) \phi(0)\>} &=& 0, \nn\\
\frac{\< \phi_H(\infty) \phi_H(0) T(w_1) T(w_2) \>}{\< \phi(\infty) \phi(0)\>} &=& c\frac{z_1 z_2 }{\alpha^4 w_1^2 w_2^2} \left[ \frac{z_1 z_2}{2z_{12}^4} - \frac{(\alpha^2-1)}{12 z_{12}^2}\right].
\label{eq:phiphiTT}
\ee
Expanding the above $\< \phi_H \phi_H T T\>$ correlator at $w_1 \sim w_2$, one can see that there is only a fourth-order pole at $w_1 \sim w_2$ and the higher order poles cancel, as is enforced by the $TT$ OPE. 
To compute the $1/c$ correction to the leading order heavy-light Virasoro blocks, the only modes we need to sum are single- and double-$\CL$ modes. Calculating the overlap factors with the light operators and the inner product factors that enter is a straightforward application of the Virasoro algebra.  It will be convenient to use a basis of double-$L$ modes that are symmetric in the indices, i.e. of the form $L_{(m,n)} \equiv \frac{L_{m}L_{n} + L_{n}L_{m}}{2}$. One finds
\be
\< L_{(m,n)} \phi_L(z) \phi_L(0) \> &=& \frac{1}{2} h_L \left(2 (m-1) (n-1) h_L+(m-1) m+(n-1) n\right)z^{m+n-2h_L}, \nn\\
\< L_{m+n} \phi_L(0) \phi_L(0)\> &=& h_L(m+n-1)z^{m+n-2h_L},
\ee
and
\be
{\cal M}_{(m,n), (m,n) } &\equiv& \< L_{(m,n)} L_{(-m,-n)} \> , \nn\\
{\cal M}_{(m,n), (m+n)} &\equiv & \< L_{(m,n)} L_{-(m+n)} \>, \nn\\
{\cal M}_{(m+n), (m+n)} &\equiv &  \< L_{m+n} L_{-(m+n)} \>.
\ee
 Inverting and expanding to $\CO(1/c^2)$,  
\be
\CM^{-1}_{(m,n), (m,n)} &=& \frac{144}{c^2 n m (n^2-1)(m^2-1)(1+\delta_{n,m})} + \CO(1/c^3), \\
\CM^{-1}_{(m,n),(a)} &=& -72\frac{ \frac{(m+2 n)}{n \left(n^2-1\right) } + \frac{(n+2 m)}{m \left(m^2-1\right) } }{c^2 (m+n-1) (m+n) (m+n+1)(1+\delta_{n,m})}\delta_{a,m+n} + \CO(1/c^3). \nn
\ee
Now, these factors can be substituted into the sum that defines the projector.  We can take advantage of the fact that $\< \phi_H \phi_H T(w_1) T(w_2)\>$ is a generating function for $\< \phi_H \phi_H \CL_{-n} \CL_{-m}\>$ in order to write these terms as contour integrals in the following form: 
\be
\< \phi_H \phi_H {\cal P}_{h,w} \phi_L \phi_L \> &=&\sum_{m,n} \< \phi_H \phi_H \CL_{-n} \CL_{-m}\> \left( \CM^{-1}_{(m,n), (m,n)} \< \CL_m \CL_n \phi_L \phi_L\>+ \CM^{-1}_{(m,n), (m+n)} \< \CL_{m+n}\phi_L \phi_L\>  \right) \nn\\
 && =   \left( \frac{w'(z)w'(1)}{w^2} \right)^{h_L}\oint \frac{dw_1}{2 \pi i w_1} \frac{dw_2}{2 \pi i w_2} \< \phi_H \phi_H T(w_1) T(w_2) \> G(w_1, w_2), 
 \label{eq:ContourProj}
\ee
where
\be
G(w_1,w_2) &=& \left( \frac{w'(z)w'(1)}{w^2} \right)^{-h_L}\sum_{m,n}\frac{w_1^2w_2^2}{w_1^nw_2^m} \left(\CM^{-1}_{(m,n), (m,n)} \< \CL_m \CL_n \phi_L \phi_L\>+ \CM^{-1}_{(m,n), (m+n)} \< \CL_{m+n}\phi_L \phi_L\>  \right) \nn\\
 &=& \sum_{n\ge 2, m\ge 2} \left( \frac{w}{w_1} \right)^n \left( \frac{w}{w_2} \right)^m \frac{72 h_L w_2^2 w_1^2 \left(h_L (m+n) (m+n+1)+m n\right)}{c^2 m (m+1) n (n+1) (m+n) (m+n+1)} . 
\label{eq:SCGreen}
\ee
The sum on $m$ in $G(w_1, w_2)$ can be done in closed form, and we get a combination of powers, logs, and hypergeometrics of the form
\be
{}_2F_1 \left( 1,2+n,4+n,\frac{w}{w_2} \right) , \qquad {}_2F_1 \left( 1,3+n,4+n,\frac{w}{w_2} \right).
\ee
   The integration contour in (\ref{eq:ContourProj}) must have $|w|< |w_1| < 1, |w|< |w_2|<1$, since the sum over powers of $w_1, w_2$ converges in $G_2$ when $|w|< |w_1|, |w_2|$, and the sum over powers of $w_1, w_2$ in $\< \phi \phi TT\>$ converges when $1> |w_1|, |w_2|$.  Starting with the contour integral over $w_2$, we can shrink it down as far as possible.  However, the sum over $m$ produces branch cuts that prevent one from shrinking the contour all the way down to the origin.  These branch cuts in $w_2$ are along the real axis between 0 and $w$; the discontinuities across this branch cut can be read off from the coefficients of the logarithms in $G(w_1,w_2)$, together with the following expressions for the discontinuities of the hypergeometric functions:
 \be
{\rm disc}_{{\rm Im}(w_2)\rightarrow 0}\left[ {}_2F_1 \left(1,2+n,4+n,\frac{w}{w_2} \right) \right] &=& 2\pi i \frac{(2+n)(3+n) w_2^2 (w_2 - w) }{w^3} \left(\frac{w_2}{w}\right)^n\nn\\
{\rm disc}_{{\rm Im}(w_2) \rightarrow 0}\left[ {}_2F_1 \left(1,3+n,4+n, \frac{w}{w_2} \right) \right] &=& - 2 \pi i (3+n) \left(\frac{w_2}{w}\right)^{n+3}.
\ee
We therefore reduce to
\be
\< \phi_H \phi_H {\cal P}_{h,w} \phi_L \phi_L \>  &\equiv & \left( \frac{w'(z)w'(1)}{w^2} \right)^{h_L} \left( h_L {\cal V}_{h_L}^{(1)} + h_L^2 {\cal V}_{h_L^2}^{(1)}\right) , \nn\\
{\cal V}_{h_L}^{(1)} &=& w^{-2h_L}\oint \frac{dw_1}{2 \pi i w_1} \int_0^w dw_2  \< \phi_H \phi_H T(w_1) T(w_2) \>  \nn\\
 && \times \sum_{n\ge 2} \frac{72 w_2 w_1^{2-n} w^{n-1} \left(\left(\frac{w_2}{w}\right){}^n \left(n w_2-n w-w_2 \right) + w_2\right)}{c n
   \left(n^2-1\right)}, \nn\\
{\cal V}_{h_L^2}^{(1)} &=&  w^{-2h_L}\oint \frac{dw_1}{2 \pi i w_1} \int_0^w dw_2  \< \phi_H \phi_H T(w_1) T(w_2) \>  \nn\\
 && \times\sum_{n \ge 2} \frac{72 w_2 w_1^{2-n} w^{n-1} \left(w- w_2 \right)}{c n (n+1)}  . 
\ee
We interpret the ${\cal V}_{h_L}^{(1)}$ term as a true `quantum' correction while ${\cal V}_{h_L^2}^{(1)}$ is `semi-classical'.  The former would correspond to a loop effect in AdS$_3$, while the latter captures effects from classical gravitational backreaction from the light probe object.\footnote{Note that there is no $\CO(h_L^0)$ piece.  In fact, this is true at all orders in $1/c$, since such a term would have to survive in the limit that $h_L =0$. But in that case, $\phi_L$ would have to be the identity operator, so the vacuum ``block'' would be the $\<\phi_H(\infty) \phi_H(1)\>$ two-point function, which is just  constant normalized to 1. } 
Finally, the remaining sum on $n$ converges in the region that $|w_1|>|w_2|$, and gives 
\be
{\cal V}_{h_L}^{(1)} &=& \frac{w^{-2h_L}}{c w^2}\oint \frac{dw_1}{2 \pi i w_1} \int_0^w dw_2  \< \phi_H \phi_H T(w_1) T(w_2) \>  \nn\\
 && \times\left( -36 w_1 \left(w_2^2  \left(w_1-w\right){}^2 \log \left(1-\frac{w}{w_1}\right)  \right. \right. \nn\\
  && \left. \left. +\left(w_2-w_1\right) w \left(w_2
   \left(w_2-w\right)+\left(2 w_1 w_2-\left(w_1+w_2\right) w\right) \log \left(1-\frac{w_2}{w_1}\right)\right)\right) \right), \nn\\
{\cal V}_{h_L^2}^{(1)} &=&  w^{-2h_L}\oint \frac{dw_1}{2 \pi i w_1} \int_0^w dw_2  \< \phi_H \phi_H T(w_1) T(w_2) \>  \nn\\
 && \times\frac{36 w_1 w_2 \left(w-w_2\right) \left(w \left(2 w_1-w\right)+2 w_1 \left(w_1-w\right) \log
   \left(1-\frac{w}{w_1}\right)\right)}{c w^2}. 
\ee
  Thus, we can shrink the $w_1$ contour onto the branch cut from 0 to $w$. However, note that after we do this, the  branch cut from $\log(1-\frac{w_2}{w_1})$ is crossed when  $w_1<w_2$, but not when $w_1>w_2$.  One also crosses a pole at $w_1 \sim w_2$ in $\< \phi_H \phi_H T T\>$.  However, as explained below equation (\ref{eq:phiphiTT}), the only such singularity is $\< \phi_H \phi_H TT\> \sim \frac{c/2}{w_{12}^4}$. This does not contribute to any $\< \phi_H \phi_H \CL_{-n} \CL_{-m}\>$ overlap term with $n, m \ge 2$, since in a small $w_1, w_2$ expansion it does not have any terms with non-negative powers of both $w_1$ and $w_2$, so we can just subtract it out.   Taking this into account, we finally obtain
   \be
   {\cal V}_{h_L}^{(1)} &=& -\frac{36 }{c w^2} \int_0^w dw_1 \int_0^w dw_2 \< \phi_H(\infty) \phi_H(1) T(w_1)T(w_2)\>' \nn\\
   && \times \left[ w_2^2 \left(w_1-w\right){}^2 - \Theta(w_2 - w_1) \left(w_1-w_2\right) w \left(2 w_1 w_2-\left(w_1+w_2\right) w\right) \right] \nn\\
   {\cal V}_{h_L^2}^{(1)} &=& -\frac{72 }{c w^2} \int_0^w dw_1 \int_0^w dw_2 \< \phi_H(\infty) \phi_H(1) T(w_1)T(w_2)\>' \nn\\
   && \times \left[ w_1 w_2 \left(w_1-w\right)\left(w_2-w\right) \right] 
   \label{eq:FinalInt}
   \ee
   The primes on the correlators indicate that we are to subtract out their $\sim \frac{c/2}{w_{12}^4}$ singularities.
 
 The function ${\cal V}_{h_L^2}^{(1)}$ contributes to the conformal block at $\CO(\frac{h_L^2}{c})$ times a function of $\alpha$, and consequently it is part of the ``semi-classical'' piece.  The semi-classical part is defined as the piece of $\log {\cal V}$ that is formally of $\CO(c)$ in the limit where $c$ is large and $h_H/c, h_L/c$ are held fixed.   However, the function ${\cal V}_{h_L}^{(1)}$ contributes only to $\log {\cal V}$ at $\CO(c^0)$ in this limit and therefore goes beyond the semi-classical part of the block.
 
We were able to evaluate both the semi-classical and quantum $1/c$ corrections, which are written closed form in equation \ref{eq:ExactResult}.  In what follows we will examine some interesting limits of the general result.

\subsection{Small $h_H$ limit}

The main reason that the integrals in (\ref{eq:FinalInt}) are difficult is that $\< \phi_H(\infty) \phi_H(1) T(w_1) T(w_2)\>$ written as a function of $w_1,w_2$ contains non-integer powers of $w_1,w_2$ arising from $z_i = 1-(1-w_i)^{1/\alpha}$.  In the limit that $h_H/c$ is small, we can expand the correlator $\< \phi_H \phi_H T T\>$ around $\alpha=1$, and these become integer powers and logarithms.  At $\CO(\alpha-1)$, one has 
\be
&&\< \phi_H(\infty) \phi_H(1) T(w_1) T(w_2) \>' \nn\\
 &&= (\alpha-1)\frac{c \left(-\frac{\left(w_1^2+2 \left(5 w_2-6\right) w_1+\left(w_2-12\right)
   w_2+12\right) w_{12}}{\left(w_1-1\right) \left(w_2-1\right)}-6
   \left(w_1+w_2-2\right) \log \left(\frac{1-w_2}{1-w_1}\right)\right)}{6 w_{12}^5} \nn\\
    && + \CO((\alpha-1)^2).
   \ee
   
The resulting integrals in (\ref{eq:FinalInt}) can be easily evaluated.  The result is 
\begin{equation}
\begin{aligned}
  \CV_{h_L^2}^{(1)} &= 6  \frac{(\alpha-1)}{z^2} \left(4 z^2+2 (z-1) \log ^2(1-z)-(z-2) z \log (1-z)\right)   + \CO \left( (\alpha-1)^2 \right)  \\
   {\cal V}_{h_L}^{(1)} &= \frac{ (\alpha-1)}{z^2}  \left(-6 (z-2) z
   \left(\text{Li}_2\left(\frac{1}{1-z}\right)+\text{Li}_2(z)\right)+\left(\pi
   ^2 (z-2)-16 z\right) z\right. 
\\
   & \left. -3 (3 (z-2) z+2) \log ^2(1-z)+(z-2) z (6 \log (z)+6 i
   \pi -1) \log (1-z)\right) \\ 
    +&  \CO((\alpha-1)^2) .
   \end{aligned}
   \label{eq:VLinAlphaLinH}
   \end{equation} 
Since $w = z + \CO(\alpha-1)$, there is no difference between using $w$ vs $z$ in the expression at leading order in $\CO(\alpha-1)$  above.  We have checked these expressions against a direct small $z$ expansion up to $\CO(z^8)$ using the methods of \cite{ZamolodchikovRecursion}. 

\subsection{Large $T$ limit}

It is more interesting to consider limits that allow $\alpha = 2 \pi i T_L$ to be imaginary, since that is the regime where the heavy state develops a horizon in AdS and a temperature.  The limit that is most likely to be generic is that where $T_L$ is taken to $\infty$.  In particular, as mentioned in the introduction, in this limit one can rescale distance as $x\rightarrow x/T$ to obtain the infinite radius limit of the circle.  While the two-point function on the circle at finite radius and finite temperature is equivalent to a two-point function on the torus and is thus not a universal quantity, the two-point function on the plane at finite temperature {\it is} the universal function (\ref{eq:ThermalEqualsPure1}), independent of all CFT data except for the dimension $h_L$.  

Fortunately, $T \rightarrow \infty$ is also a limit where the integrand (\ref{eq:FinalInt}) simplifies significantly:
\be
&&\< \phi_H(\infty) \phi_H(1) T(w_1) T(w_2) \>' \nn\\
&& = \frac{1}{12} c \left(\frac{6-\log
   ^2\left(\frac{1-w_2}{1-w_1}\right)}{\left(w_1-1\right){}^2 \left(w_2-1\right){}^2 \log
   ^4\left(\frac{1-w_2}{1-w_1}\right)}-\frac{6}{\left(w_1-w_2\right){}^4}\right) .
\ee
Substituting this into (\ref{eq:FinalInt}), we obtain the result 
   \be
   {\cal V}_{h_L}^{(1)} &=&  -\frac{1}{c w^2} \left[ 24 (w-1)^2 \text{Ei}(-\log (1-w))+24 \text{li}(1-w)+10 w^2-24 \gamma_E  ((w-2) w+2) \right. \nn\\
     && \left. +48 (w-1)
   \log (-\log (1-w))-w ((26-25 w) \log (1-w)+24 w \log (w))\right] \nn\\
 {\cal V}_{h_L^2}^{(1)}&=& -\frac{12 }{c w^2} \left[ -4 (w-1) \text{Ei}(-\log (1-w))-4 (w-1) \text{li}(1-w)-2 w^2-w^2 \log \left((1-w) \log ^2(1-w)\right)\right.\nn\\
  && \left. +2 w^2 \log (w)-8 i \pi  w+8 \gamma_E  w+8
   (w-1) \log (\log (1-w))+8 i \pi -8 \gamma_E\right] ,
   \label{eq:InfTBlock}
   \ee
   where Ei and li  are the exponential and  logarithm integral functions, respectively.

Since the periodicity in Eulidean time is expected to be $1/T$, in the infinite temperature limit we want to scale $t$ to zero with $t T$ fixed.  The variable $w$ depends on $t$ through $w=1-e^{2 \pi i T t}$, so whether or not the block is periodic in $t T$ is a question of it monodromy as $w$ is taken around 1 in the complex plane.  One can start by looking at the behavior of (\ref{eq:InfTBlock}) around $w \sim 1$:
\be
{\cal V}_{h_L^2}^{(1)} &=& 
-\frac{12}{c} \left( \log(1-w) +2 + 2 \log(-\log(1-w)) + \dots \right) ,
  \nn\\
{\cal V}_{h_L}^{(1)} &=& \frac{1}{c} \left( \log(1-w) +24\gamma_E -10 + \dots \right) .
\ee 
The presence of these logarithms lead to non-trivial monodromies around $w =1$, and as a result the vacuum block on its own is not periodic in time.\footnote{Since $1-w=e^{2 \pi i T t}$ has unit norm, it is necessary to check the monodromy not just in a small $1-w$ expansion.  This is straightforward to do using (\ref{eq:InfTBlock}) and does indeed contain a non-trivial monodromy as $T t \rightarrow Tt + 1$.}

 \subsection{Dependence on $T$}
 \label{sec:DependenceonT}
 
 Next, we want to consider how the $1/c$ correction varies as a function of temperature.  Note that the first several terms in the small $w$ expansion at large $\alpha$ and at small $\alpha$ are remarkably similar:
 \be
 {\cal V}_{h_L}^{(1)} &=&\left\{ \begin{array}{cc} 
   - \frac{(\alpha-1)  }{75 c} w^4 \left( 1 + 2w + 2.781 w^2 + 3.342 w^3 + 3.728 w^4 + \dots \right) \qquad \alpha \rightarrow 1
  \\
 - \frac{11 }{1800 c} w^4 \left( 1+ 2w + 2.783 w^2+ 3.350 w^3+ 3.744 w^4 + \dots \right) \qquad \alpha \rightarrow i \infty
 \end{array} \right. \nn\\
   {\cal V}_{h_L^2}^{(1)} &=& \left\{ \begin{array}{cc} 
   - \frac{(\alpha-1)}{15 c} w^4 \left( 1+2w + 2.786 w^2 + 3.357 w^3 + 3.757 w^4 + \dots \right) \qquad \alpha \rightarrow 1
    \\
   - \frac{11}{360 c} w^4 \left( 1 + 2w+ 2.793 w^2 + 3.380 w^3 + 3.800 w^4 + \dots \right) \qquad \alpha \rightarrow i \infty
    \end{array} \right. 
  \ee
The fact that both begin as $1 + 2 w$ follows from global conformal symmetry, but the similarity of the subsequent terms is non-trivial. It reflects the fact that each additional `graviton' is making a suppressed contribution, so that both functions are well-approximated at small $w$ by the lowest dimension 2-graviton global block $w^4 {}_2 F_1(4,4,8,w)$.  As discussed in the next section, we believe  that the similarity at $\alpha = 1$ and $\alpha = i \infty$ is a consequence of the fact that the BTZ solution is simply an orbifold of AdS$_3$, although it would be interesting to see this explicitly.

\begin{figure}[ht]
\includegraphics[width=0.5\textwidth]{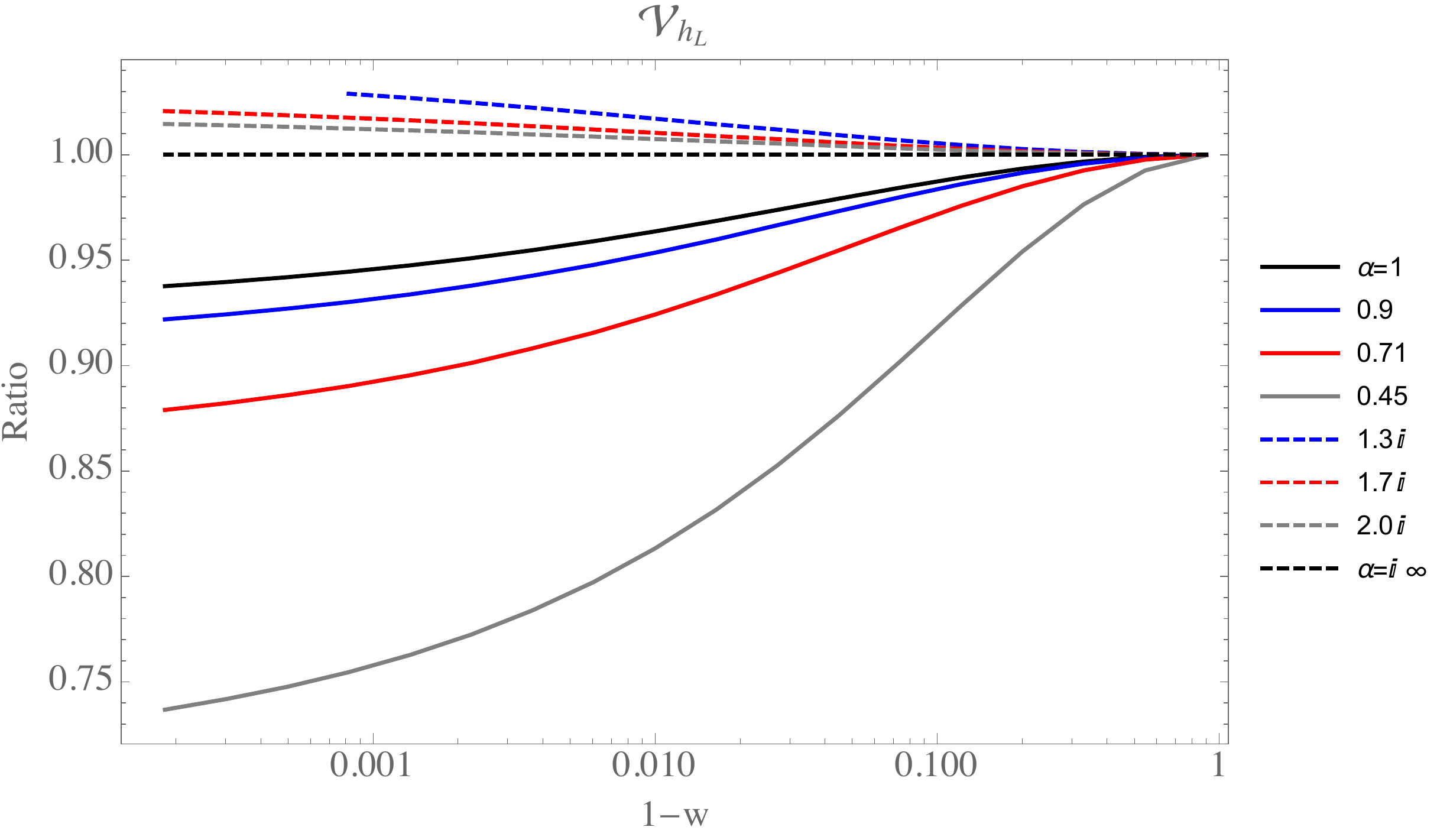}
\includegraphics[width=0.5\textwidth]{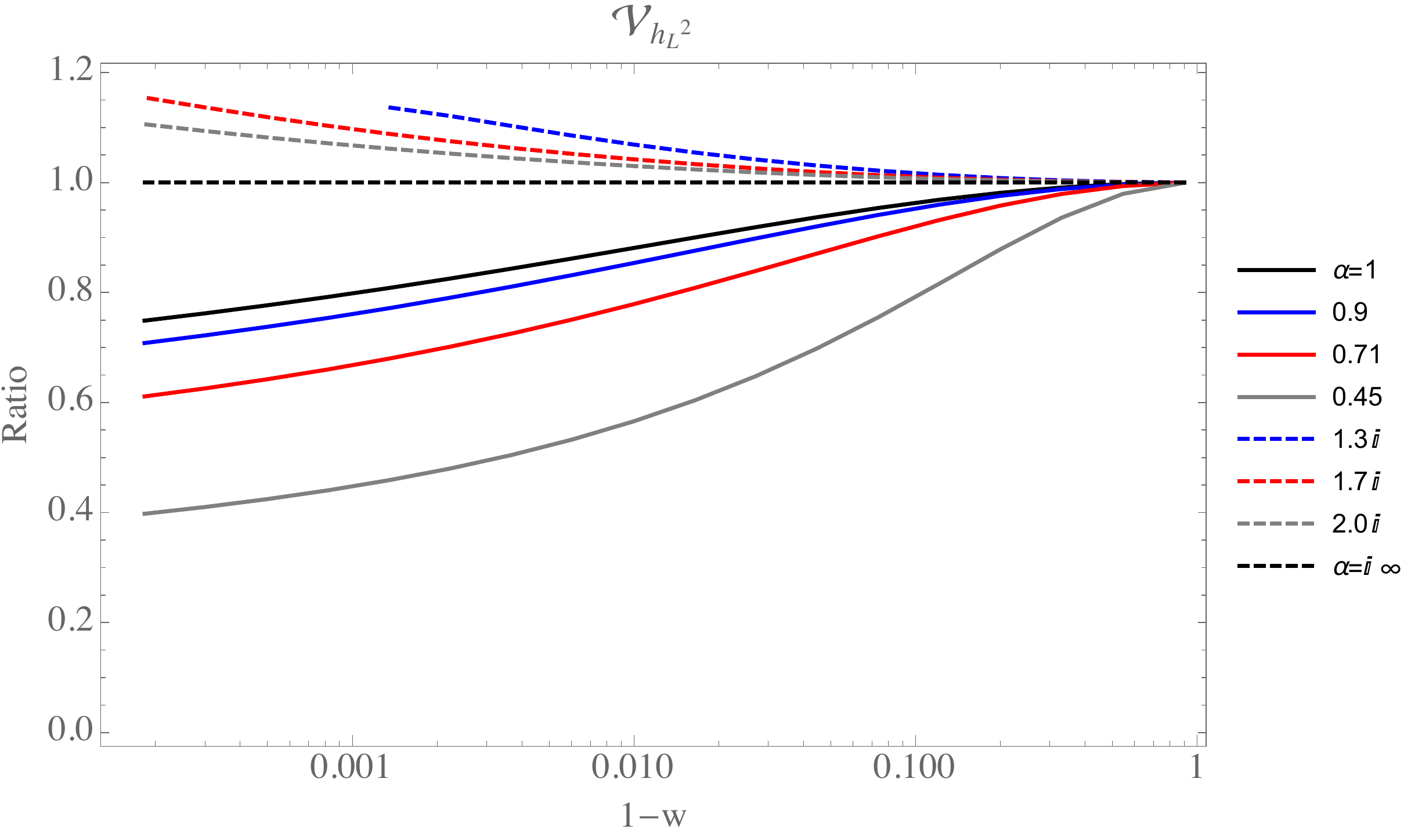}

\includegraphics[width=0.5\textwidth]{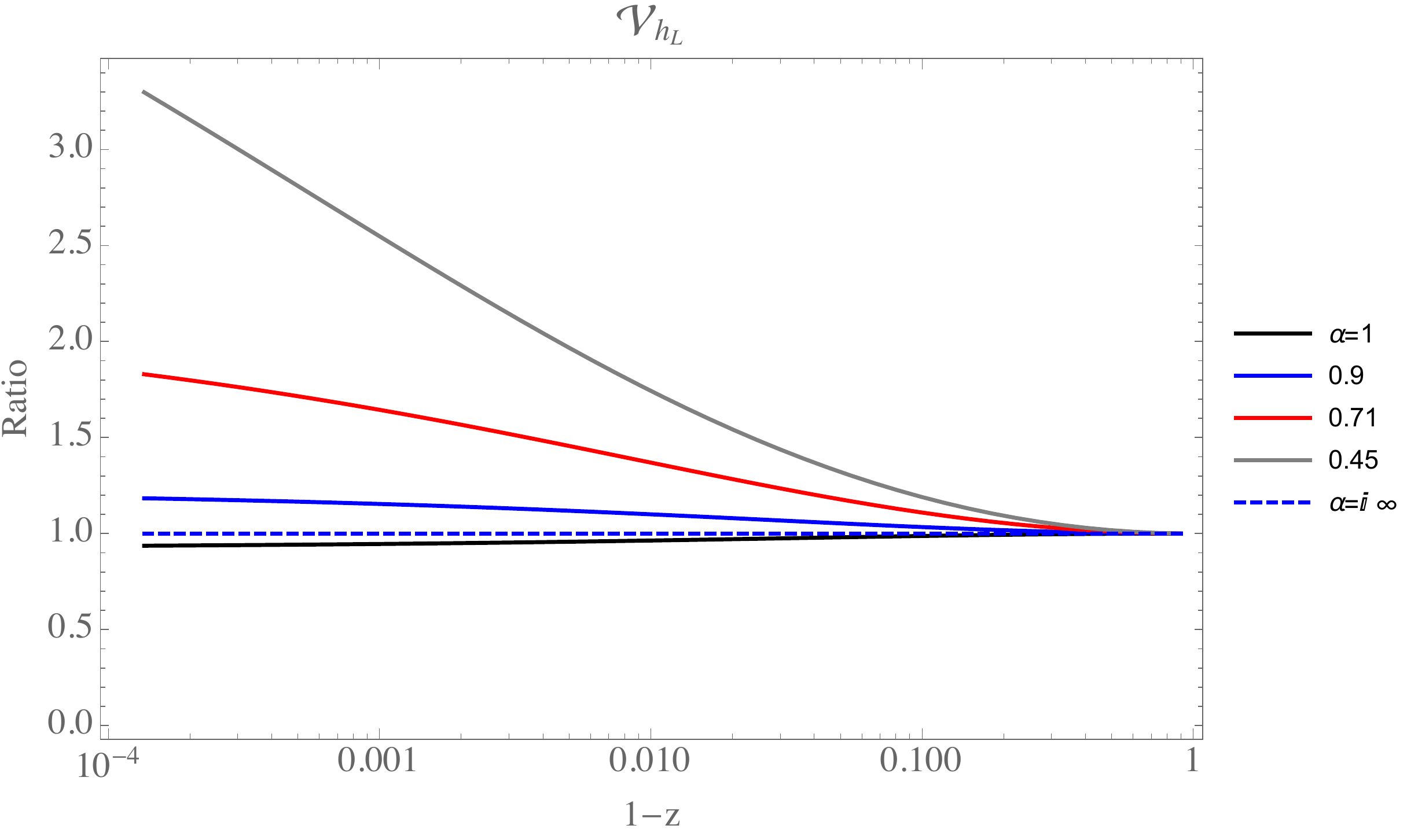}
\includegraphics[width=0.5\textwidth]{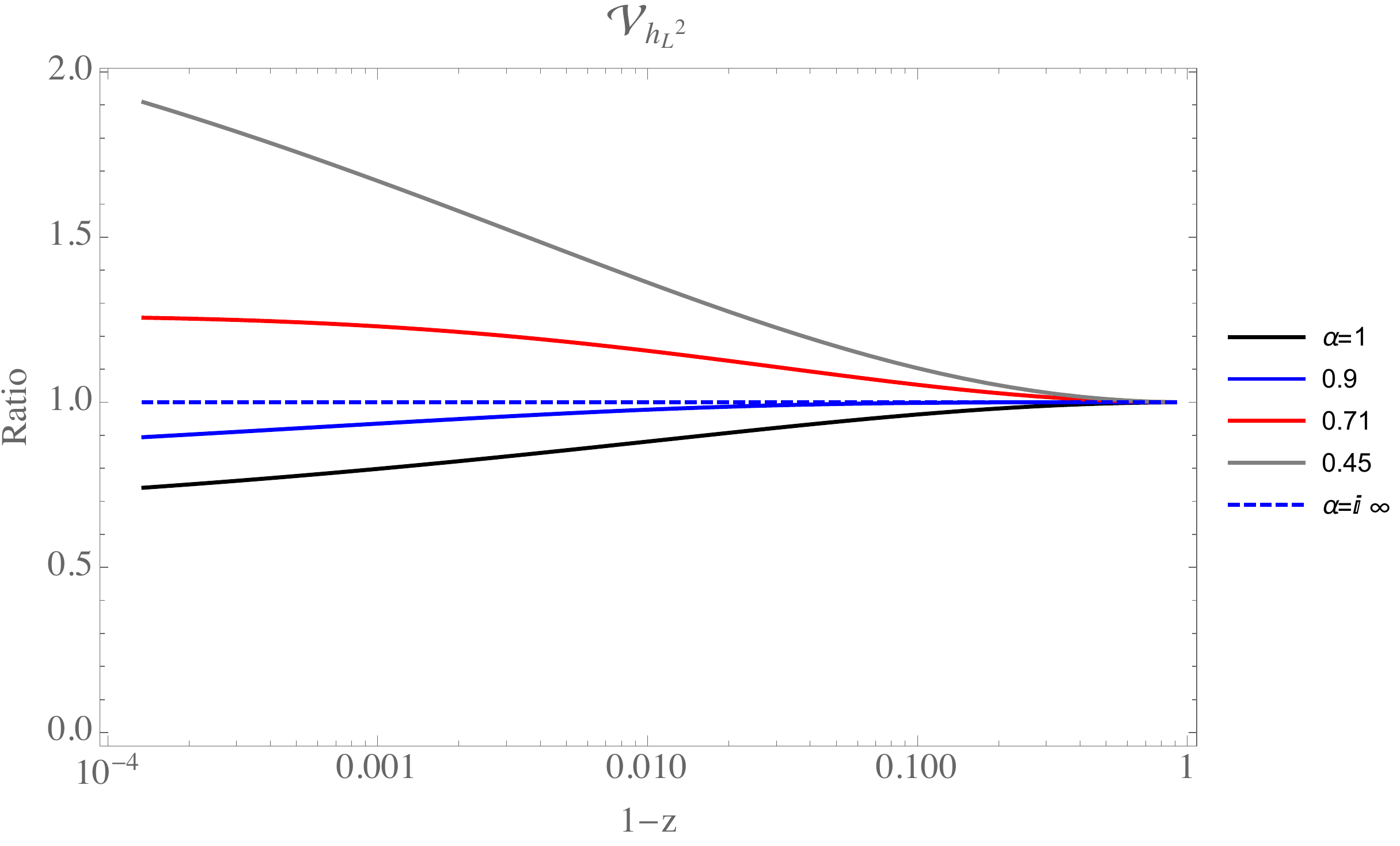}
\caption{This figure shows the similarity in the functional dependence of ${\cal V}_{h_L}^{(1)}$ and ${\cal V}_{h_L^2}^{(1)}$ for different values of $\alpha$.  {\it Left,top:} The ratio $N(\alpha) \frac{{\cal V}^{(1)}_{h_L}(w; \alpha)}{{\cal V}^{(1)}_{h_L}(w ;\alpha=i \infty)}$, where a normalization $N(\alpha)= -\frac{11 \alpha ^4}{(1-\alpha ) (\alpha +1) \left(11 \alpha ^2+1\right)}$ scales them to agree at $z \sim 0$.  {\it Right,top:} Same as the left, but for the $\CO(h_L^2)$ term $N(\alpha) \frac{{\cal V}^{(1)}_{h_L^2}(w; \alpha)}{{\cal V}^{(1)}_{h_L^2}(w; \alpha=i \infty)}$. The endpoints $\alpha=1$ and $\alpha = i \infty$ are very close, but the difference becomes more significant near $\alpha \sim 0$. {\it Left and right, bottom}: Same as the top, but as a function of $z$ for real $z$.}
\label{fig:alphacomp}
\end{figure}

 In Figure \ref{fig:alphacomp} we plot the $w$-dependence for various values of $\alpha$ to show this agreement explicitly.  As one can see there, it is only near $\alpha \sim 0$ (which is the minimum threshold for black holes in AdS$_3$) that the $w$-dependence differs significantly from either the $\alpha = 1$ or $\alpha = i \infty$ extreme.    

\section{Expectations from Thermodynamics and AdS/CFT}
\label{sec:ExpectationsThermoAdSCFT}

In the last section we computed perturbative $1/c$ corrections to the Virasoro conformal blocks in the heavy-light limit.  Unlike the leading order vacuum block, these corrections appear to deviate from expectations from thermodynamics, or equivalently, from black hole physics in AdS$_3$, as they have non-trivial monodromies in Euclidean time.  In what follows we will explain this in more detail, and then show that our results do not necessarily differ from expectations from AdS/CFT.  The main point is that individual conformal blocks generically have unphysical monodromies that can cancel when they are summed to compute full CFT correlators, and that these monodromies have a simple origin in AdS.

\subsection{Periodicity in Euclidean Time and Pure State Thermodynamics}
\label{sec:KMS}

Let us summarize the well-known features of field theory correlation functions in the canonical ensemble, to facilitate comparison with the pure state correlation functions and associated conformal blocks that we have studied.\footnote{For a more general discussion see e.g. section 4.1.2 of \cite{Papadodimas:2012aq}.  The connection between Lorentzian and Euclidean correlators in a CFT context was extensively reviewed in \cite{Hartman:2015lfa}.}
 
The thermal 2-pt function is
\be
F_{12}(t_L, \vec x) \equiv \tr  \left( e^{- \beta H} \CO_1(t_L, \vec x) \CO_2(0) \right)  
\ee
where we emphasize that $t_L$ is a Lorentzian time coordinate.  Inserting a complete set of states shows
\be
F_{12}(t_L, \vec x) &=& \sum_{\psi, \psi'} \< \psi | e^{- (\beta - i t_L) H} \CO_1(0, \vec x) e^{ -i t_L H} | \psi' \> \< \psi'| \CO_2(0) | \psi \> 
\nn \\
&=& \sum_{\psi, \psi'} \< \psi'| e^{ i t_L H} \CO_2(0) e^{- (\beta + i t_L) H} | \psi \>  \< \psi | \CO_1(0, \vec x)   | \psi' \> 
\ee
which leads to the KMS condition
\be
F_{12}(t_L - i \beta, \vec x) = F_{21}(t_L, \vec x)
\ee
stating that the correlator is periodic in imaginary time, up to an exchange of the order of the operators.
In relativistic QFTs, the two operators commute at space-like separations $|t_L| < | \vec x|$, which means that $F_{12}$ and $F_{21}$ must be analytic continuations of each other.  From the single Euclidean correlator $\mathcal{F}(t_E, \vec x)$ we can obtain either $F_{12}$ or $F_{21}$ by approaching the lightcone branch cuts of $\mathcal{F}$ at $t_L^2 = \vec x^2$ from different sides.  For the cases that we will be studying $\CO_1 = \CO_2 = \CO_L$.

In recent work \cite{Fitzpatrick:2014vua, Fitzpatrick:2015zha, Fitzpatrick:2015foa} we have been comparing thermal 2-pt correlators with the 4-pt correlator in a heavy background
\be \label{eq:ThermalEqualsPure}
 \< \CO_H(\infty) \CO_L(1) \CO_L(z) \CO_H(0) \> \approx \< \CO_L(1) \CO_L(z) \>_{T} = \left( \frac{\pi T}{\sin(\pi T t)} \right)^{2h_L} ,
\ee
where $z = 1 - e^{-t + i \phi}$.  In CFT$_2$ the thermal 2-pt correlator of Virasoro primary operators is uniquely fixed via a conformal mapping from the plane to the cylinder.  The thermal correlator agrees precisely with the large $c$ heavy-light Virasoro vacuum conformal block, where $2 \pi  T = \sqrt{\frac{24 h_H}{c} - 1}$ is the temperature.\footnote{Both sides of the identity can accomodate separate holomorphic and anti-holomorphic temperatures $T$ and $\bar T$, with the case $T \neq \bar T$ corresponding to a  spinning BTZ black hole in AdS/CFT.}   

The limit of large central charge with fixed $h_H / c$ can be interpreted as a high-energy limit in a theory with many-degrees of freedom.  Thus we expect an identity such as equation (\ref{eq:ThermalEqualsPure}), because in the thermodynamic limit, a pure state drawn from the canonical (or micro-canonical) ensemble should be very difficult to distinguish from the true thermal density matrix.  In AdS/CFT, this is the statement that black holes and very high energy microstates should be nearly identical.  In fact, in AdS$_3$ there are no approximately stable orbits around black holes, so these states are even more `inescapable' than in higher dimensions.  We expect that order-by-order in the $1/c$ expansion, heavy-light correlators will appear thermal, and that only non-perturbatively small effects may violate the approximate KMS condition in heavy-light correlators.

We  pause to note a subtlety concerning the identification in equation (\ref{eq:ThermalEqualsPure}):
we should really be comparing the full heavy-light 4-pt correlator with $\< \CO_L \CO_L \>_T$ on the torus, since both functions must be periodic in the angular $\phi$ coordinate under $\phi \to \phi + 2 \pi$.  But the 2-pt function on the torus is not fixed by conformal invariance;  this corresponds to the fact that Virasoro blocks other than the vacuum will contribute to the complete heavy-light 4-pt function.  The vacuum does make an important universal contribution, but for example from an AdS$_3$ description there would also be double-trace $\CO_L \partial^n \CO_L$ contributions that sum up to restore the perioidicity in $\phi$ at any $t$.  One can avoid these complications by studying the light-cone OPE limit \cite{Fitzpatrick:2012yx, KomargodskiZhiboedov}, or by taking the limit of $T \to \infty$ with $Tt$ fixed, so that the $\phi$ direction is effectively non-compact when distances are measured in units of $1/T$.  In that large temperature limit and at large $c$, the identification of equation (\ref{eq:ThermalEqualsPure}) becomes precise.

In section \ref{sec:CorrectionstoBlock} we computed the $1/c$ corrections to the heavy-light Virasoro conformal blocks and found deviations from the thermal result that could not be interpreted as a perturbative renormalization of the temperature.  In the next sections we will discuss how conformal blocks can be computed from AdS in order to explain why our thermality-violating $1/c$ corrections should not necessarily be interpreted as a violation of the Euclidean-time-periodicity seen in equation (\ref{eq:ThermalEqualsPure}).  To be precise, we need to distinguish between two different notions of thermality.  The first, which is specific to 2d CFTs, is that at infinite $T$ (or equivalently, through rescaling, in a CFT in non-compact space), the two-point function should be {\it exactly} (\ref{eq:ThermalEqualsPure}).  The second is that the two-point function should be periodic in Euclidean time.  Knowledge of the vacuum Virasoro block is sufficient to see that the first of these is violated, assuming even a mild $\CO(1)$ gap in dimensions of operators.  The reason is that (\ref{eq:ThermalEqualsPure}) makes a definite prediction for the coefficients of OPE singularities in the four-point correlator, and low-order singularities can receive contributions only from low-dimension operators.  Thus, a small gap is enough to imply that the first few such singularities receive contributions from only the vacuum block, and therefore that the OPE does not match the prediction of (\ref{eq:ThermalEqualsPure}).   This is in contrast with the second, more general, criterion for thermality, which requires knowledge of the correlator at finite values of $t$ and therefore depends on  the full operator content of the theory; this will be the main focus of the following sections. 
  However, our results do show that in the lightcone OPE limit, where the Virasoro vacuum block dominates (assuming no additional conserved currents), the form of the $1/c$ corrections imply that the correlator cannot be separately periodic in $t \pm i \phi$.

As a final comment, note that we can reproduce the exact canonical ensemble by summing over individual pure microstates, so that
\be
 \< \CO_L(1) \CO_L(z) \>_{T} \equiv \sum_{\CO}  e^{-E_\CO / T} \< \CO(\infty) \CO_L(1) \CO_L(z) \CO(0) \> 
\ee
In this relation we must let the sum range over both Virasoro primaries and descendants, whereas in equation (\ref{eq:ThermalEqualsPure}) we have been focusing on Virasoro primaries $\CO_H$.  In a CFT$_2$ where $\< \CO_L(1) \CO_L(z) \>_{T}$ is entirely fixed by conformal invariance, this relation provides a constraint on CFT data closely related to modular invariance.

\subsection{Monodromies of Global Conformal Blocks from AdS}
\label{sec:MonodromyGlobalBlocks}

Local operators in QFTs commute at spacelike separation, so CFT correlators like
\be
\< \CO_H(x_1) \CO_H(x_2) \CO_L(x_3) \CO_L(x_4) \>
\ee
are single valued analytic functions of the Euclidean $x_i$, with singularities only occurring in the OPE limits where $x_i$ and $x_j$ coincide.  This property also holds when  CFT correlators are obtained from a quantum field theory in AdS via the AdS/CFT dictionary and the bulk Feynman diagram expansion.  

However, conformal blocks do not have this property.  For example, consider a conformal block in the channel $HH \to LL$, which can be computed as a sum over intermediate states
\be
G_{\Delta, \ell} = \< \CO_H(\infty) \CO_H(1)  \left( \sum_{\alpha \, \mathrm{desc} \, \CO_{\Delta, \ell} } | \alpha \> \< \alpha | \right) \CO_L(z,\bar z) \CO_L(0) \>
\ee
where the states $| \alpha \>$ are all in the irreducible representation of a primary state/operator $\CO_{\Delta, \ell}$ with dimension and total angular momentum $\Delta, \ell$.  Equivalently, this can be computed by expanding in the OPE limit $z, \bar z \to 0$.

\begin{figure}[t!]
\begin{center}
\includegraphics[width=0.7\textwidth]{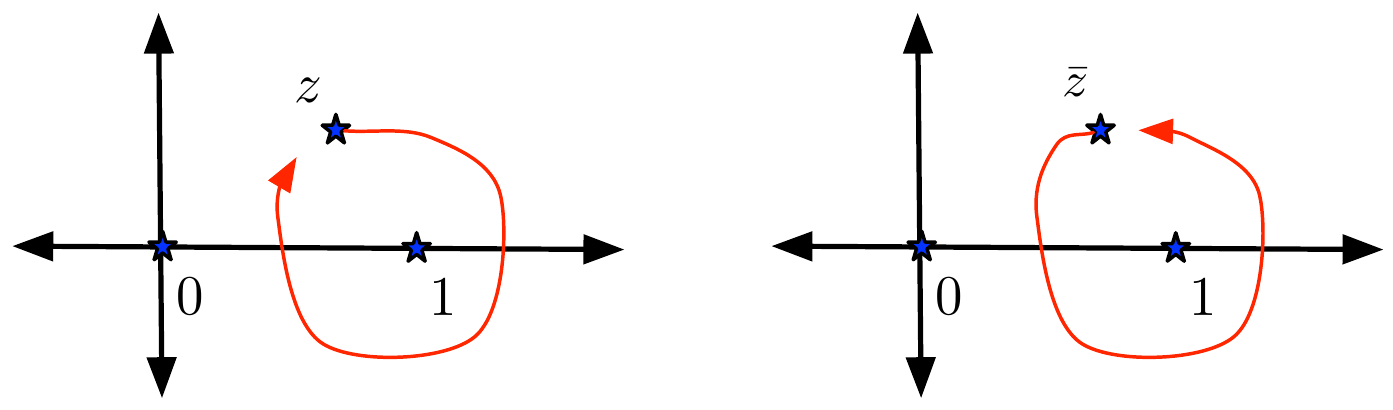}
\caption{ This figure shows the analytic continuation in $z$ and $\bar z$ that are equivalent to rotating $\CO_L(z)$ around the global AdS cylinder.  An operator at infinity is not displayed.}
\label{fig:AnalyticContinuationPhi} 
\end{center}
\end{figure}

Analytically continuing the variables $z$ and $\bar z = z^*$ around the heavy operator $\CO_H(1)$, as pictured in figure \ref{fig:AnalyticContinuationPhi}, we venture outside regime where the $\CO_L(z, \bar z) \CO_L(0)$ OPE converges.  If we interpret $1 - z = e^{-t + i \phi}$ as a coordinate on the cylinder (so that the positions of $\CO_H(1)$ and $\CO_L(0)$ are effectively switched), then we are simply continuing $\phi \to \phi + 2 \pi$.  To be very explicit, in the case of global 2d conformal blocks we can write
\be
G_{h, \bar h}(z, \bar z) = F_h(z) F_{\bar h} (\bar z) + F_h(\bar z) F_{\bar h} (z), \ \ \mathrm{with} \ \ F_\beta(x) \equiv x^\beta {}_2 F_1(\beta, \beta, 2 \beta, x)
\ee
where $\Delta = h + \bar h$ and $\ell = | h - \bar h |$.
The hypergeometric functions have logarithmic branch cuts around $z, \bar z = 1$ with non-trivial monodromies.

We would now like to explain how these monodromies arise from an AdS calculation.  First of all, note that since we are studying conformal blocks, not CFT correlators, we are not asking a question about standard bulk Feynman diagrams.  These diagrams must be single-valued in the Euclidean region.

\begin{figure}[t!]
\begin{center}
\includegraphics[width=0.35\textwidth]{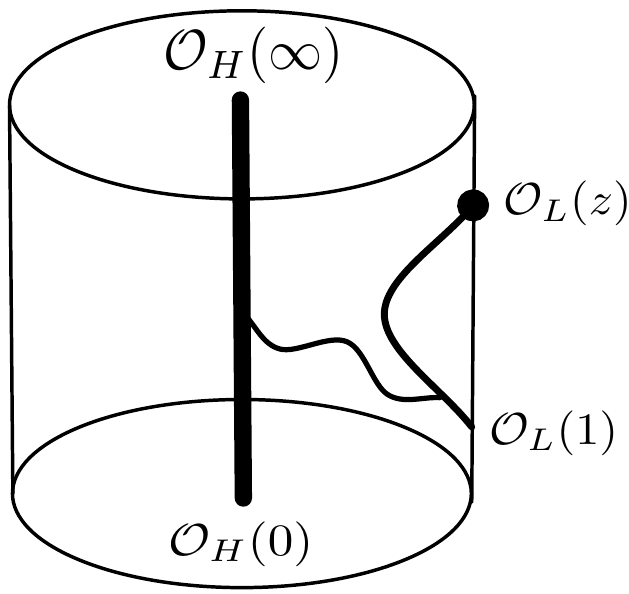}
\caption{ This figure depicts a `geodesic Witten diagrams' that can be used to compute a conformal block from AdS \cite{Hijano:2015zsa}.  The lines connecting the two light operators to each other and the two heavy operators to each other are both geodesics, while the wavy line designates a propagator whose endpoints have been fixed to these geodesics.  The only integrals are over the positions of the bulk-to-bulk propagator along the geodesics.}
 \label{fig:DiagramGeodesicConformalBlock} 
\end{center}
\end{figure}

However, as has been shown recently \cite{Hijano:2015zsa}, both global and Virasoro conformal blocks \cite{Hijano:2015qja} can be computed from a certain simplified version of a bulk Feynman diagram, which the authors of \cite{Hijano:2015zsa} refer to as `geodesic Witten diagrams'.  To obtain a geodesic Witten diagram, we begin with a Feynman diagram for a 4-pt CFT correlator with four boundary-to-bulk propagators and a single bulk-to-bulk exchange propagator $G_{BB}(X,Y)$, as pictured in figure \ref{fig:DiagramGeodesicConformalBlock}.  But instead of allowing $X$ and $Y$ to range over AdS, we confine these bulk points to geodesics when computing the diagram.  The geodesics always connect the pairs of operators whose OPE limits define the conformal block.

We can give a simple heuristic explanation of the origin of geodesic Witten diagrams as follows; for more rigorous derivations see \cite{Hijano:2015zsa}.  A 4-pt tree-level Witten diagram computation in AdS can always be decomposed \cite{Liu} (see \cite{Hijano:2015zsa, AdSfromCFT} for recent discussions) into one `single-trace' conformal block and an infinite sum of `double-trace' conformal blocks, where the former corresponds to the state exchanged in the bulk-to-bulk propagator, and the latter correspond to the external states.  In the limit that the external states have very large dimension, the bulk computation will be well-approximated by a geodesic Witten diagram via the geometric optics approximation for the heavy bulk states; in the same limit, the double-trace contributions decouple.  Thus in general we expect that the unique `single-trace' conformal block must correspond to the geodesic Witten diagram.  

\begin{figure}[t!]
\begin{center}
\includegraphics[width=0.9\textwidth]{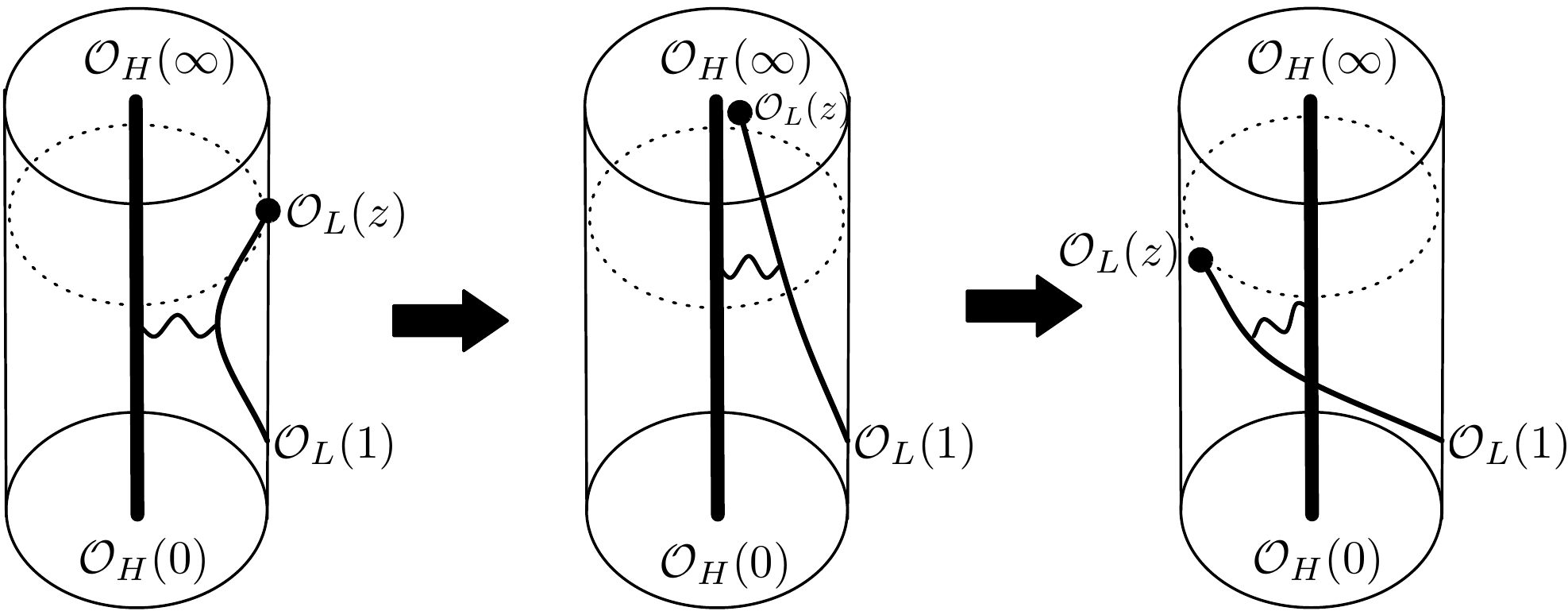}
\caption{ This figure shows what happens when we analytically continue the external points of a geodesic Witten diagram.  As $z$ moves around the cylinder, the heavy and light geodesics must cross, and as they do, the propagator connecting them passes through its short-distance singularity.  Note that in $d>2$ dimensions this crossing is enforced by geometry, not by topology. This is the origin of the non-trivial monodromy of the conformal block. Similar reasoning leads to a monodromy in Euclidean time for non-vacuum heavy-light Virasoro blocks \cite{Fitzpatrick:2015zha}.}
 \label{fig:DiagramGeodesicMonodromy} 
\end{center}
\end{figure}

Given that conformal blocks can be computed as geodesic Witten diagrams, it is easy to discover the AdS origin of their non-trivial monodromies.  The analytic continuation of figure \ref{fig:AnalyticContinuationPhi} can be applied to a geodesic Witten diagram computation, which takes the schematic form
\be
G_{\Delta,0}(z, \bar z) &=& \int_{-\infty}^{\infty} d \lambda d \lambda' G_{b \partial}(-\infty, X(\lambda)) G_{b \partial}(\infty, X(\lambda))
\nn \\
&& \times \left( \frac{e^{-2 \Delta \sigma(X, Y)}}{e^{-2  \sigma(X, Y)} - 1} \right) G_{b \partial}(1, Y(\lambda')) G_{b \partial} (z, Y(\lambda'))
\ee
and is pictured in figure \ref{fig:DiagramGeodesicMonodromy}.   The bulk variables $X(\lambda)$ and $Y(\lambda')$ run along the two geodesics, which are parameterized using $\lambda, \lambda'$.  The expression in parentheses is the (scalar) bulk-to-bulk propagator, with $\sigma(X, Y)$ the distance between the two bulk points.  Crucially, as $1 - z = e^{-t + i\phi}$ is continued in $\phi$, we necessarily pass through a configuration where the two geodesics cross, which requires that we integrate over the short-distance singularity of the bulk-to-bulk propagator.  Note that the pure vacuum conformal block has a trivial monodromy, since the relevant computation would not include a bulk-to-bulk propagator.

Informally, we might say that the geodesic Witten diagram treats the external operators as classical sources in the bulk, which `remember' their relative orientation.  When we compute standard Witten diagrams, the external operators are treated as quantum fields in AdS.  The path integral sums agnostically over all their bulk trajectories, destroying any `memory' of the classical trajectories.  Cancellations between the monodromies of `single-trace' and `double-trace' operators encode the eradication of this classical memory.  

In summary, individual conformal blocks have unphysical monodromies in $\phi$, even though the blocks have been computed from a physical process transpiring in a spacetime that is manifestly periodic under $\phi \to \phi + 2 \pi$.  Next let us consider an analogous question concerning thermal periodicities and Virasoro blocks.

\subsection{Monodromies of Virasoro Conformal Blocks and AdS/CFT}

Thermal states in CFT$_2$ are dual to BTZ black holes in AdS$_3$.  As discussed in section \ref{sec:KMS}, a simple way to recognize the temperature is from the Euclidean-time periodicity of the 2-pt correlator.  This feature can be observed directly in the spinless Euclidean BTZ metric
\be
ds^2 = (r^2 + \alpha^2) dt^2 + \frac{dr^2}{r^2 + \alpha^2} + r^2 d \phi^2
\ee
where $\alpha^2 \leq 1$, and $\alpha$ is imaginary in the BTZ case.
The Euclidean time coordinate must be periodically identified under $t \sim t +  1/T_H$ to avoid a singularity at the horizon $r =  |\alpha|$, where we note that the temperature is $T_H = \frac{| \alpha |}{2\pi}$.  We expect that this periodicity will be inherited by AdS/CFT correlators computed from perturbative Feynman diagrams in the black hole background.

Geodesic Witten diagrams in AdS$_3$ have  been used to obtain semi-classical Virasoro conformal blocks \cite{Hijano:2015qja}.  To leading order in the semi-classical limit, we can compute the heavy-light Virasoro blocks in the same way that we obtained global conformal blocks in section \ref{sec:MonodromyGlobalBlocks}.  The difference is that we evaluate the geodesic Witten diagrams in the gravitational background of the heavy operator, instead of in pure AdS.

\begin{figure}[t!]
\begin{center}
\includegraphics[width=0.98\textwidth]{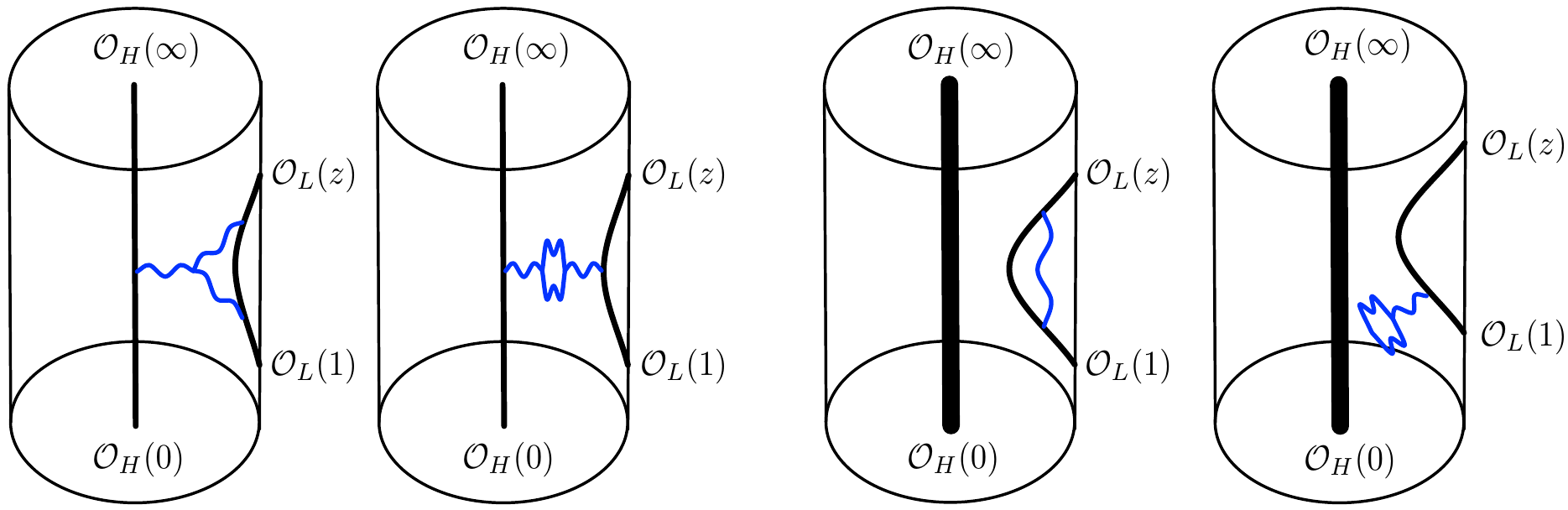}
\caption{ This figure shows gravitational one-loop diagrams in AdS that could contribute to heavy-light Virasoro blocks at order $1/c$.   In the small $h_H / c$ limit, we expect that the two diagrams on the left should correspond with the $1/c$ effects in equation (\ref{eq:VLinAlphaLinH}).
More generally, the pair of diagrams on the left should be equivalent to the pair on the right with bulk propagators computed in the background gravitational field of the heavy operator. 
}
 \label{fig:OneLoopVirasoro} 
\end{center}
\end{figure}

In the last section we studied monodromies of global conformal blocks under $\phi \to \phi + 2 \pi$.  We are now interested in Euclidean time periodicity, $t \to t + \frac{1}{T_H}$ for the Virasoro blocks.  For the case of non-vacuum blocks, the reasoning from the last section can be copied directly, replacing $\phi$ with $t$.  In fact the global AdS$_3$ metric is identical to the spinless BTZ metric in the high temperature limit, after rescaling $r \to r / r_+$ and exchanging the roles of $t$ and $\phi$.  So the monodromies of the non-vacuum heavy-light blocks first obtained in \cite{Fitzpatrick:2015zha} can be understood heuristically from the `memory' effect of the geodesic Witten diagrams \cite{Hijano:2015qja}.  

The semi-classical Virasoro vacuum block computes the exponential of a geodesic length \cite{Fitzpatrick:2014vua, HartmanExcitedStates, Hijano:2015rla} in a deficit angle or BTZ background, and in both cases it has a periodicity set by $\alpha$.  For real $\alpha$ this is a periodicity in $\phi$ associated with the deficit angle, while for imaginary $\alpha = 2 \pi i T_H$ it is periodicity in Euclidean time.   What remains is to understand 
the presence of a non-trivial monodromy in the $1/c$ correction to this vacuum block, as we found in section \ref{sec:CorrectionstoBlock}.  

The geodesic Witten diagram technology has not been applied in the presence of perturbative $1/c$ corrections, so it is not entirely clear how to proceed. Even in the case of the large $c$ semi-classical blocks, instead of full bulk propagators (which include a sum over images \cite{KeskiVakkuri:1998nw} in order to satisfy the correct boundary value problem), the authors of \cite{Hijano:2015qja} used pure AdS  propagators with a rescaling $t, \phi \to \alpha t, \alpha \phi$.  This led to the correct result, and it might be interpreted as a strategy for eliminating double-trace contributions, but it was not given an a priori derivation.  

We will proceed by discussing the most natural generalization of the geodesic Witten diagrams which leads to a single Virasoro conformal block.  Some relevant diagrams are pictured in figure \ref{fig:OneLoopVirasoro}.  The pair of diagrams on the right clearly have a different structure from those we have considered previously, and in particular, the simple reasoning of figure \ref{fig:DiagramGeodesicMonodromy} no longer applies, since there are no explicit propagators connecting the deficit angle/black hole to the light operator geodesic.  The third diagram from the left leads to an integral of the schematic form
\be
\int d \lambda_1 d \lambda_2 G_{\partial B}(1, Y_1(\lambda_1)) G_{BB}( Y_1(\lambda_1), Y_2(\lambda_2))
G_{grav}( Y_1(\lambda_1), Y_2(\lambda_2)) G_{\partial B}( Y_2(\lambda_2), z)
\ee
where $\lambda_i$ parameterize two points $Y_i(\lambda_i)$ on the light operator geodesic, and the two bulk-to-bulk propagators correspond to the light operator and the gravitational field.  

We can think of the bulk-to-bulk propagators in the BTZ background as the result of summing an infinite set of diagrams connecting a free AdS bulk-to-bulk propagator to a succession of graviton propagators.  This justifies the expectation of a non-trivial monodromy as we rotate $z$ on the thermal circle.  We will need the relevant bulk-to-bulk propagator\footnote{To perform the full computation it would probably be most expedient to use the Chern-Simons form for gravity; for a convenient form for the C-S propagator see appendix A.2 of \cite{Dong:2014tsa}.} in a deficit angle or BTZ black hole backgrounds.  Since these backgrounds are orbifolds of pure AdS$_3$, the propagators can be determined through the method of images.  This sum over images produces a new logarithmic singularity in the propagators at the location of the deficit angle and at the black hole singularity \cite{Kraus:2002iv}.  Without the sum over images the propagators have only a short-distance singularity.

Thus we are led to conjecture that the internal propagators in the diagrams on the right of figure \ref{fig:OneLoopVirasoro} should include a sum over images, so that they are sensitive to the deficit angle or black hole singularity.  The integration over such propagators could then explain the monodromy of the $1/c$ correction to the Virasoro vacuum block under analytic continuation in Euclidean time.  It would be interesting to explore this question further, and to obtain explicit agreement between our CFT$_2$ computation and a gravity calculation in AdS$_3$.

Given that we are arguing that double-trace operator conformal blocks must be included to see the correct ``thermal'' properties of the heavy-light correlator, one may wonder why the leading order in $1/c$ vacuum correlator did not suffer from non-periodic monodromies.  The simplest way to understand this is that there is a limit where the vacuum block actually is the full correlator: in the limit of infinite $T$ and infinite $c$, the contribution from  double-trace operators is indeed negligible, leaving only the vacuum block to fulfill the thermal properties of the theory.  

As a final comment, in section \ref{sec:DependenceonT} we pointed out that the functional form of the $1/c$ corrections appears very similar at large temperature and at small $h_H$, two regimes that are very different physically.  We believe that from the bulk point of view, this is due to the fact that BTZ backgrounds are locally pure AdS.  Diagrams such as those in figure \ref{fig:OneLoopVirasoro} will produce very similar corrections for all values of $\alpha$ at small $z$, where the light operator geodesics in figure \ref{fig:OneLoopVirasoro}  do not extend very far into the bulk.

\section*{Acknowledgments}

We would like to thank Brian Swingle for discussions and collaboration at various stages of this work.
We would also like to thank Tom Hartman, Simeon Hellerman, Ami Katz, Daliang Li, Eric Perlmutter, Matt Walters, and Junpu Wang for valuable discussions.  JK  is supported in part by NSF grants PHY-1316665, PHY-1454083, and by a Sloan Foundation fellowship.  ALF was partially supported by ERC grant BSMOXFORD no. 228169.  This work was performed in part at the Aspen Center for Physics, which is supported by National Science Foundation grant PHY-1066293.

\begin{appendices}

\section{Subleading Order in $1/c$ from Projectors}
\label{app:gencase}

When we expand to subleading order at large $c$ with operator dimensions held fixed, it is straightforward to see that the only states that contribute are the modes of a single stress tensor, and that the resummation of all these modes just gives the $T$ global conformal block.  We can ask whether an analogous approach is tractable in an expansion around the semi-classical limit.  Much of the structure is the same as the expansion around the classical limit.  Considering the different factors in the projector,
\be
\CP_{h,w} \approx \sum_{\{m_i,k_i\},\{m'_i,k'_i\}} \CL_{-m'_1}^{k_1} \cdots \CL_{-m'_n}^{k_n} | h \> \CM^{-1}_{\{m_i, k_i\}, \{m_i', k_i'\}}\< h_w | \CL_{m_n}^{k_n} \cdots \CL_{m_1}^{k_1}
\ee
where $\CM$ is the inner product matrix of the bra and ket states, we see that both $\CM$ and $\< h_w | \CL_{m_n}^{k_n} \dots \CL_{m_1}^{k_1} \phi_L \phi_L \>$ are exactly the same as in the classical limit, up to a conformal transformation acting on the light fields.  The only factor that changes is the left-action,
\be
\< \phi_H \phi_H  \CL_{-m'_1}^{k_1} \cdots \CL_{-m'_n}^{k_n} | h \> .
\ee
We begin by calculating the overlap with the light states. As in the body of the article, it will be convenient to work in a basis that is symmetric in the indices, $L_{(m,n)} \equiv \frac{L_m L_n + L_n L_m}{2}$.   The overlap with the light states is a straightforward exercise in commuting the $\CL$'s toward the right, to obtain
\be
&&\< h | L_{m+n} \phi_{L_1} (z) \phi_{L_2}(0)\> = C_{h\phi\phi} (h-h_2+h_1(m+n))z^{h-h_1-h_2+m+n}, \nn\\
&&\< h | L_{(m,n)} \phi_{L_1}(z) \phi_{L_2}(0)\>  \\
  &&=  C_{h\phi\phi}  \frac{ ( h-h_2 + h_1 m+n)(h-h_2 + h_1 n) + (h- h_2 +h_1 m)(h-h_2 + m + h_1 n)}{2} z^{h-h_1 -h_2+m+n}. \nn
\ee

The inner product factors are also straightforward, though require much more book-keeping.  The only non-vanishing matrix elements are of the form
\be \< h | L_m L_n L_{-n} L_{-m} | h \>, \< h | L_{m+n} L_{-n} L_{-m} | h \>, \< h | L_m L_n L_{-n-m} | h \>, \< h | L_{m+n} L_{-n-m} | h \>.
\ee Taking into account mixing between states $L_{(m,n)}$ and $L_{(a,b)}$ with $a+b = m+n$, we find
\be
\CM^{-1}_{(1),(1)} &=& \frac{1}{2h},  \\
\left( \begin{array}{cc} \CM^{-1}_{(2),(2)} & \CM^{-1}_{(2),(1,1)} \\
 \CM^{-1}_{(1,1),(2)} &  \CM^{-1}_{(1,1),(1,1)} \end{array} \right)  &=& \left( \begin{array}{cc} \frac{2}{c} + \frac{ 4(5-8h)h}{c^2 (1+2h)} & - \frac{3}{c(1+2h)} + \frac{6h(-5+8h)}{c^2(1+2h)^2} \\
- \frac{3}{c(1+2h)} + \frac{6h(-5+8h)}{c^2(1+2h)^2}  & \frac{1}{4h(1+2h)} + \frac{9}{2c (1+2h)^2} + \frac{9(5-8h)h}{c^2(1+2h)^3} \end{array} \right), \nn
\ee
For $m\ge 2, n\ge 2$, one has
\begin{equation}
\begin{aligned}
\CM^{-1}_{(m+n),(m+n)} &= \left[ 2 c^2 h (m+n)^2 \left((m+n)^2-1\right)^2 \right]^{-1} \\
 &\times 3 \Big(2 h (m+n-1) ((m+n) ((m+n) (4 c+3 (m+n-2) (m+n)+55)+4 (c+1))-12)   \\
  & +(m+n-1)
   (m+n+1)^2 (c (m+n-2) (m+n)-24)-192 h^2 (m+n)   \\
    & -48 h (m+n) \left((m+n)^2-1\Big)
   H_{m+n-2}\right), \\
   \CM^{-1}_{(m+1), (m,1)} &= -\frac{3 \left(-24 h \left(m^2-2 m-2\right)-24 m (m+2)\right)}{c^2 h (m-1) m^2 (m+1)^2
   (m+2)}-\frac{3 \left(m^2-1\right)}{c h (m-1) m (m+1)^2} ,  \\
    \CM^{-1}_{(m,1),(m,1)} &= -\frac{144 (h+1)}{c^2 h (m-1)^2 m (m+1)^2}-\frac{6 \left(1-m^2\right)}{c h (m-1)^2 m
   (m+1)^2} \\
    \CM^{-1}_{(m+n), (m,n)} &= -\frac{72 \left(m^4+2 m^3 n-m^2+2 m n \left(n^2-2\right)+n^4-n^2\right)}{c^2 m
   \left(m^2-1\right) n \left(n^2-1\right) (m+n-1) (m+n) (m+n+1) (\delta _{m,n}+1)}\\
   \CM^{-1}_{(m,n), (m,n)} &= \frac{144}{c^2 m \left(m^2-1\right) \left(n^3-n\right) (\delta _{m,n}+1)}.
   \end{aligned}
   \end{equation}
 In section \ref{sec:computation}, we combined both the inner product factors and the overlap factors with the light operators into a single function $G(z_1, z_2)$.  We can do the same thing here, except now we have two functions, one for a single $\CL$ in the overlap with the heavy operators, and one for a double $\CL$ in the overlap with the heavy operators.  That is, 
\be
G_1(w_1) &=& \left( \frac{w'(z)^{h_1} w'(1)^{h_2} }{w^{h_1 +h_2-h_{\CO}}} \right)^{-1} \sum_{a=1}^\infty \frac{w_1^2}{w_1^a} \CM^{-1}_{(a),(a)} \<h|  \CL_a \phi_L \phi_L \> \\
G_2(w_1,w_2) &=& \left( \frac{w'(z)^{h_1} w'(1)^{h_2} }{w^{h_1 +h_2-h_{\CO}}} \right)^{-1}  \nn\\
 && \times  \sum_{m,n=1 \atop m\ge n}^\infty \frac{w_1^2w_2^2}{w_1^nw_2^m} \left(\CM^{-1}_{(m,n), (m,n)} \< h| \CL_m \CL_n \phi_L \phi_L\>+ \CM^{-1}_{(m,n), (m+n)} \< h| L_{m+n}\phi_L \phi_L\>  \right). \nn
\ee

To get the functions $G_1, G_2$ as the above sum, one just needs to put together the expressions for the overlap with the light operators and the expressions for the inner product factors.  The resulting expressions are quite lengthy and so we do not present them here, since in any case it is easy to substitute the explicit expressions above for the constituent factors. %
Next, we need to evaluate the overlap with the heavy operators.  As mentioned earlier, this is most easily performed by using the fact that correlation functions with $T$ are effectively generating functions for these overlap factors.  Since $w^2 T(w)$ is holomorphic, we can compute its correlators under a conformal transformation to $w$ coordinates by using the singularities of its OPE in these coordinates, or by starting with standard Euclidean space formulae for correlators of $T(z)$s in terms of correlators without them and then explicitly performing the conformal transformation to $w$ coordinates.  In any case, we obtain
\be
\< \phi_{H_1}(\infty) \phi_{H_2}(0) T(w) \CO(z_\CO)\> &=& \frac{C_{h \phi \phi} }{\alpha^2 w^2 z_\CO^{h+h_2-h_1}} \left( \frac{h_\CO z z_\CO}{(z_\CO-z)^2} - \frac{h_{12}}{2} \frac{z_\CO+z}{z_\CO-z} \right)
\ee
for the single-$T$ correlator, and 
\be
&&\alpha^4 w_1^2 w_2^2 \< \phi_{H_1}(\infty) \phi_{H_2}(0) T(w_1) T(w_2) \CO(1)\> \nn\\
&& \qquad = c \left( \frac{ z_1^2 z_2^2}{2 z_{12}^4}-\frac{\left(\alpha ^2-1\right)  z_1 z_2}{12 z_{12}^2} \right) 
\nn\\
 && \qquad
  + \frac{h_\CO^2}{\left(z_1-1\right){}^2 \left(z_2-1\right){}^2}+\frac{\left(h_\CO+h_{12}\right) h \left(z_1
   z_2-1\right)}{\left(z_1-1\right){}^2 \left(z_2-1\right){}^2} \nn\\
    && \qquad +\frac{h_{12} \left(h_{12} \left(z_1-1\right)+h z_1\right)}{2
   \left(z_1-1\right){}^2}+\frac{h_{12} \left(h_{12} \left(z_2-1\right)+h z_2\right)}{2 \left(z_2-1\right){}^2}+\frac{h_{12}
   \left(h_\CO+h_{12}\right)}{\left(z_1-1\right) \left(z_2-1\right)} \nn\\
    && \qquad +\frac{z_1 z_2 \left(h_\CO \left(z_1+z_2\right)+h_{12} \left(z_1
   z_2-1\right)\right)}{\left(z_1-1\right) \left(z_2-1\right) z_{12}^2}+\frac{h_{12}^2}{4}
   \ee
   for the double-$T$ correlator.

The conformal block in terms of these factors is given by the following integral:
\be
\< \phi_H \phi_H {\cal P}_{h,w} \phi_L \phi_L \> &=&  \left( \frac{w'(z)^{h_1} w'(1)^{h_2} }{w^{h_1 +h_2-h_{\CO}}} \right) \nn\\
 && \times \Big[ \oint \frac{dw_1}{2 \pi i w_1} \frac{dw_2}{2 \pi i w_2} \< \phi_{H_1}(\infty) \phi_{H_2}(1) T(w_1) T(w_2) \CO(0) \> G_2(w_1, w_2) \nn\\
&& +  \oint \frac{dw_1}{2 \pi i w_1} \< \phi_{H_1}(\infty) \phi_{H_2}(1) T(w_1)\CO(0) \> G_1(w_1) \Big].
\ee

\end{appendices}

\newpage

\bibliographystyle{utphys}
\bibliography{VirasoroBib}

%bibliography generated by nb.bst v1.01 (C) 2003-2010 Niklas Beisert
\begin{thebibliography}{10}
\ifx\href\asklfhas\newcommand{\href}[2]{#2}\fi
\ifx\arxivref\asklfhas\newcommand{\arxivref}[2]{\href{http://arxiv.org/abs/#1}{#2}}\fi
\ifx\doiref\asklfhas\newcommand{\doiref}[2]{\href{http://dx.doi.org/#1}{#2}}\fi
\parskip 0pt
\normalsize

\bibitem{Cardy:1986ie}
J.~L. Cardy,
\textit{``{Operator Content of Two-Dimensional Conformally Invariant
  Theories}''},
\doiref{10.1016/0550-3213(86)90552-3}{Nucl.Phys. \textbf{B270}, 186 (1986)}.
%%CITATION = NUPHA,B270,186;%%

\bibitem{Hartman:2014oaa}
T.~Hartman, C.~A. Keller \& B.~Stoica,
\textit{``{Universal Spectrum of 2d Conformal Field Theory in the Large c
  Limit}''},
\doiref{10.1007/JHEP09(2014)118}{JHEP \textbf{1409}, 118 (2014)},
\normalsize{\texttt{\arxivref{1405.5137}{arXiv:1405.5137}}}.
%%CITATION = ARXIV:1405.5137;%%

\bibitem{Strominger:1996sh}
A.~Strominger \& C.~Vafa,
\textit{``{Microscopic origin of the Bekenstein-Hawking entropy}''},
\doiref{10.1016/0370-2693(96)00345-0}{Phys.~Lett. \textbf{B379}, 99 (1996)},
\normalsize{\texttt{\arxivref{hep-th/9601029}{hep-th/9601029}}}.
%%CITATION = HEP-TH/9601029;%%

\bibitem{Strominger:1997eq}
A.~Strominger,
\textit{``{Black hole entropy from near horizon microstates}''},
\doiref{10.1088/1126-6708/1998/02/009}{JHEP \textbf{9802}, 009 (1998)},
\normalsize{\texttt{\arxivref{hep-th/9712251}{hep-th/9712251}}}.
%%CITATION = HEP-TH/9712251;%%

\bibitem{Maldacena:1997re}
J.~M. Maldacena,
\textit{``{The Large N limit of superconformal field theories and
  supergravity}''},
\doiref{10.1023/A:1026654312961, 10.1023/A:1026654312961}{Adv.Theor.Math.Phys.
  \textbf{2}, 231 (1998)},
\normalsize{\texttt{\arxivref{hep-th/9711200}{hep-th/9711200}}}.

\bibitem{Witten}
E.~Witten,
\textit{``{Anti-de Sitter space and holography}''},
Adv.~Theor.~Math.~Phys. \textbf{2}, 253 (1998),
\normalsize{\texttt{\arxivref{hep-th/9802150}{hep-th/9802150}}}.
%%CITATION = HEP-TH/9802150;%%

\bibitem{GKP}
S.~S. Gubser, I.~R. Klebanov \& A.~M. Polyakov,
\textit{``{Gauge theory correlators from non-critical string theory}''},
\doiref{10.1016/S0370-2693(98)00377-3}{Phys.~Lett. \textbf{B428}, 105 (1998)},
\normalsize{\texttt{\arxivref{hep-th/9802109}{hep-th/9802109}}}.
%%CITATION = HEP-TH/9802109;%%

\bibitem{Fitzpatrick:2014vua}
A.~L. Fitzpatrick, J.~Kaplan \& M.~T. Walters,
\textit{``{Universality of Long-Distance AdS Physics from the CFT
  Bootstrap}''},
\doiref{10.1007/JHEP08(2014)145}{JHEP \textbf{1408}, 145 (2014)},
\normalsize{\texttt{\arxivref{1403.6829}{arXiv:1403.6829}}}.
%%CITATION = ARXIV:1403.6829;%%

\bibitem{Fitzpatrick:2012yx}
A.~L. Fitzpatrick, J.~Kaplan, D.~Poland \& D.~Simmons-Duffin,
\textit{``{The Analytic Bootstrap and AdS Superhorizon Locality}''},
\doiref{10.1007/JHEP12(2013)004}{JHEP \textbf{1312}, 004 (2013)},
\normalsize{\texttt{\arxivref{1212.3616}{arXiv:1212.3616}}}.
%%CITATION = ARXIV:1212.3616;%%

\bibitem{KomargodskiZhiboedov}
Z.~Komargodski \& A.~Zhiboedov,
\textit{``{Convexity and Liberation at Large Spin}''},
\doiref{10.1007/JHEP11(2013)140}{JHEP \textbf{1311}, 140 (2013)},
\normalsize{\texttt{\arxivref{1212.4103}{arXiv:1212.4103}}}.
%%CITATION = ARXIV:1212.4103;%%

\bibitem{Shenker:2013pqa}
S.~H. Shenker \& D.~Stanford,
\textit{``{Black holes and the butterfly effect}''},
\doiref{10.1007/JHEP03(2014)067}{JHEP \textbf{1403}, 067 (2014)},
\normalsize{\texttt{\arxivref{1306.0622}{arXiv:1306.0622}}}.
%%CITATION = ARXIV:1306.0622;%%

\bibitem{Roberts:2014ifa}
D.~A. Roberts \& D.~Stanford,
\textit{``{Two-dimensional conformal field theory and the butterfly effect}''},
\normalsize{\texttt{\arxivref{1412.5123}{arXiv:1412.5123}}}.
%%CITATION = ARXIV:1412.5123;%%

\bibitem{Shenker:2014cwa}
S.~H. Shenker \& D.~Stanford,
\textit{``{Stringy effects in scrambling}''},
\doiref{10.1007/JHEP05(2015)132}{JHEP \textbf{1505}, 132 (2015)},
\normalsize{\texttt{\arxivref{1412.6087}{arXiv:1412.6087}}}.
%%CITATION = ARXIV:1412.6087;%%

\bibitem{Maldacena:2015waa}
J.~Maldacena, S.~H. Shenker \& D.~Stanford,
\textit{``{A bound on chaos}''},
\normalsize{\texttt{\arxivref{1503.01409}{arXiv:1503.01409}}}.
%%CITATION = ARXIV:1503.01409;%%

\bibitem{Mathur:2009hf}
S.~D. Mathur,
\textit{``{The Information paradox: A Pedagogical introduction}''},
\doiref{10.1088/0264-9381/26/22/224001}{Class.~Quant.~Grav. \textbf{26}, 224001
  (2009)},
\normalsize{\texttt{\arxivref{0909.1038}{arXiv:0909.1038}}},
in \textit{``{Strings, Supergravity and Gauge Theories. Proceedings, CERN
  Winter School, CERN, Geneva, Switzerland, February 9-13 2009}''},
p.~224001.
%%CITATION = ARXIV:0909.1038;%%

\bibitem{Mathur:2010kx}
S.~D. Mathur,
\textit{``{The Information paradox and the infall problem}''},
\doiref{10.1088/0264-9381/28/12/125010}{Class.~Quant.~Grav. \textbf{28}, 125010
  (2011)},
\normalsize{\texttt{\arxivref{1012.2101}{arXiv:1012.2101}}}.
%%CITATION = ARXIV:1012.2101;%%

\bibitem{Almheiri:2012rt}
A.~Almheiri, D.~Marolf, J.~Polchinski \& J.~Sully,
\textit{``{Black Holes: Complementarity or Firewalls?}''},
\normalsize{\texttt{\arxivref{1207.3123}{arXiv:1207.3123}}}.
%%CITATION = ARXIV:1207.3123;%%

\bibitem{Braunstein:2009my}
S.~L. Braunstein, S.~Pirandola \& K.~Zyczkowski,
\textit{``{Better Late than Never: Information Retrieval from Black Holes}''},
\doiref{10.1103/PhysRevLett.110.101301}{Phys.~Rev.~Lett. \textbf{110}, 101301
  (2013)},
\normalsize{\texttt{\arxivref{0907.1190}{arXiv:0907.1190}}}.
%%CITATION = ARXIV:0907.1190;%%

\bibitem{Papadodimas:2012aq}
K.~Papadodimas \& S.~Raju,
\textit{``{An Infalling Observer in AdS/CFT}''},
\doiref{10.1007/JHEP10(2013)212}{JHEP \textbf{1310}, 212 (2013)},
\normalsize{\texttt{\arxivref{1211.6767}{arXiv:1211.6767}}}.
%%CITATION = ARXIV:1211.6767;%%

\bibitem{Papadodimas:2013jku}
K.~Papadodimas \& S.~Raju,
\textit{``{State-Dependent Bulk-Boundary Maps and Black Hole
  Complementarity}''},
\doiref{10.1103/PhysRevD.89.086010}{Phys.~Rev. \textbf{D89}, 086010 (2014)},
\normalsize{\texttt{\arxivref{1310.6335}{arXiv:1310.6335}}}.
%%CITATION = ARXIV:1310.6335;%%

\bibitem{Harlow:2014yoa}
D.~Harlow,
\textit{``{Aspects of the Papadodimas-Raju Proposal for the Black Hole
  Interior}''},
\doiref{10.1007/JHEP11(2014)055}{JHEP \textbf{1411}, 055 (2014)},
\normalsize{\texttt{\arxivref{1405.1995}{arXiv:1405.1995}}}.
%%CITATION = ARXIV:1405.1995;%%

\bibitem{Maldacena:2001kr}
J.~M. Maldacena,
\textit{``{Eternal black holes in anti-de Sitter}''},
\doiref{10.1088/1126-6708/2003/04/021}{JHEP \textbf{0304}, 021 (2003)},
\normalsize{\texttt{\arxivref{hep-th/0106112}{hep-th/0106112}}}.
%%CITATION = HEP-TH/0106112;%%

\bibitem{OPEConvergence}
D.~Pappadopulo, S.~Rychkov, J.~Espin \& R.~Rattazzi,
\textit{``{OPE Convergence in Conformal Field Theory}''},
\doiref{10.1103/PhysRevD.86.105043}{Phys.Rev. \textbf{D86}, 105043 (2012)},
\normalsize{\texttt{\arxivref{1208.6449}{arXiv:1208.6449}}}.
%%CITATION = ARXIV:1208.6449;%%

\bibitem{GGP}
M.~Gary, S.~B. Giddings \& J.~Penedones,
\textit{``{Local bulk S-matrix elements and CFT singularities}''},
\doiref{10.1103/PhysRevD.80.085005}{Phys.~Rev. \textbf{D80}, 085005 (2009)},
\normalsize{\texttt{\arxivref{0903.4437}{arXiv:0903.4437}}}.
%%CITATION = 0903.4437;%%

\bibitem{JP}
I.~Heemskerk, J.~Penedones, J.~Polchinski \& J.~Sully,
\textit{``{Holography from Conformal Field Theory}''},
\doiref{10.1088/1126-6708/2009/10/079}{JHEP \textbf{0910}, 079 (2009)},
\normalsize{\texttt{\arxivref{0907.0151}{arXiv:0907.0151}}}.
%%CITATION = 0907.0151;%%

\bibitem{Maldacena:2015iua}
J.~Maldacena, D.~Simmons-Duffin \& A.~Zhiboedov,
\textit{``{Looking for a bulk point}''},
\normalsize{\texttt{\arxivref{1509.03612}{arXiv:1509.03612}}}.
%%CITATION = ARXIV:1509.03612;%%

\bibitem{Hijano:2015zsa}
E.~Hijano, P.~Kraus, E.~Perlmutter \& R.~Snively,
\textit{``{Witten Diagrams Revisited: The AdS Geometry of Conformal Blocks}''},
\normalsize{\texttt{\arxivref{1508.00501}{arXiv:1508.00501}}}.
%%CITATION = ARXIV:1508.00501;%%

\bibitem{Hijano:2015qja}
E.~Hijano, P.~Kraus, E.~Perlmutter \& R.~Snively,
\textit{``{Semiclassical Virasoro Blocks from AdS$_3$ Gravity}''},
\normalsize{\texttt{\arxivref{1508.04987}{arXiv:1508.04987}}}.
%%CITATION = ARXIV:1508.04987;%%

\bibitem{Hijano:2015rla}
E.~Hijano, P.~Kraus \& R.~Snively,
\textit{``{Worldline approach to semi-classical conformal blocks}''},
\doiref{10.1007/JHEP07(2015)131}{JHEP \textbf{1507}, 131 (2015)},
\normalsize{\texttt{\arxivref{1501.02260}{arXiv:1501.02260}}}.
%%CITATION = ARXIV:1501.02260;%%

\bibitem{Beccaria}
M.~Beccaria, A.~Fachechi \& G.~Macorini,
\textit{``{Virasoro vacuum block at next-to-leading order in the heavy-light
  limit}''},
\normalsize{\texttt{\arxivref{1511.05452}{arXiv:1511.05452}}}.
%%CITATION = ARXIV:1511.05452;%%

\bibitem{ZamolodchikovRecursion}
A.~Zamolodchikov,
\textit{``{Conformal Symmetry in Two-Dimensions: An Explicit Recurrence Formula
  for the Conformal Partial Wave Amplitude}''},
\doiref{10.1007/BF01214585}{Commun.Math.Phys. \textbf{96}, 419 (1984)}.
%%CITATION = CMPHA,96,419;%%

\bibitem{Zamolodchikovq}
A.~Zamolodchikov,
\textit{``{Conformal Symmetry in Two-dimensional Space: Recursion
  Representation of the Conformal Block}''},
Teoreticheskaya~i~Matematicheskaya~Fizika \textbf{73}, 103 (1987).
%%CITATION = CMPHA,96,419;%%

\bibitem{HartmanLargeC}
T.~Hartman,
\textit{``{Entanglement Entropy at Large Central Charge}''},
\normalsize{\texttt{\arxivref{1303.6955}{arXiv:1303.6955}}}.
%%CITATION = ARXIV:1303.6955;%%

\bibitem{HarlowLiouville}
D.~Harlow, J.~Maltz \& E.~Witten,
\textit{``{Analytic Continuation of Liouville Theory}''},
\doiref{10.1007/JHEP12(2011)071}{JHEP \textbf{1112}, 071 (2011)},
\normalsize{\texttt{\arxivref{1108.4417}{arXiv:1108.4417}}}.
%%CITATION = ARXIV:1108.4417;%%

\bibitem{Alday:2009aq}
L.~F. Alday, D.~Gaiotto \& Y.~Tachikawa,
\textit{``{Liouville Correlation Functions from Four-dimensional Gauge
  Theories}''},
\doiref{10.1007/s11005-010-0369-5}{Lett.~Math.~Phys. \textbf{91}, 167 (2010)},
\normalsize{\texttt{\arxivref{0906.3219}{arXiv:0906.3219}}}.
%%CITATION = ARXIV:0906.3219;%%

\bibitem{Perlmutter:2015iya}
E.~Perlmutter,
\textit{``{Virasoro conformal blocks in closed form}''},
\doiref{10.1007/JHEP08(2015)088}{JHEP \textbf{1508}, 088 (2015)},
\normalsize{\texttt{\arxivref{1502.07742}{arXiv:1502.07742}}}.
%%CITATION = ARXIV:1502.07742;%%

\bibitem{Headrick}
M.~Headrick,
\textit{``{Entanglement Renyi entropies in holographic theories}''},
\doiref{10.1103/PhysRevD.82.126010}{Phys.Rev. \textbf{D82}, 126010 (2010)},
\normalsize{\texttt{\arxivref{1006.0047}{arXiv:1006.0047}}}.
%%CITATION = ARXIV:1006.0047;%%

\bibitem{Fitzpatrick:2015zha}
A.~L. Fitzpatrick, J.~Kaplan \& M.~T. Walters,
\textit{``{Virasoro Conformal Blocks and Thermality from Classical Background
  Fields}''},
\normalsize{\texttt{\arxivref{1501.05315}{arXiv:1501.05315}}}.
%%CITATION = ARXIV:1501.05315;%%

\bibitem{Fitzpatrick:2015foa}
A.~L. Fitzpatrick, J.~Kaplan, M.~T. Walters \& J.~Wang,
\textit{``{Hawking from Catalan}''},
\normalsize{\texttt{\arxivref{1510.00014}{arXiv:1510.00014}}}.
%%CITATION = ARXIV:1510.00014;%%

\bibitem{KrausBlocks}
E.~Hijano, P.~Kraus \& R.~Snively,
\textit{``{Worldline approach to semi-classical conformal blocks}''},
\normalsize{\texttt{\arxivref{1501.02260}{arXiv:1501.02260}}}.
%%CITATION = ARXIV:1501.02260;%%

\bibitem{Alkalaev:2015wia}
K.~B. Alkalaev \& V.~A. Belavin,
\textit{``{Classical conformal blocks via AdS/CFT correspondence}''},
\doiref{10.1007/JHEP08(2015)049}{JHEP \textbf{1508}, 049 (2015)},
\normalsize{\texttt{\arxivref{1504.05943}{arXiv:1504.05943}}}.
%%CITATION = ARXIV:1504.05943;%%

\bibitem{Alkalaev:2015lca}
K.~B. Alkalaev \& V.~A. Belavin,
\textit{``{Monodromic vs geodesic computation of Virasoro classical conformal
  blocks}''},
\normalsize{\texttt{\arxivref{1510.06685}{arXiv:1510.06685}}}.
%%CITATION = ARXIV:1510.06685;%%

\bibitem{Hartman:2015lfa}
T.~Hartman, S.~Jain \& S.~Kundu,
\textit{``{Causality Constraints in Conformal Field Theory}''},
\normalsize{\texttt{\arxivref{1509.00014}{arXiv:1509.00014}}}.
%%CITATION = ARXIV:1509.00014;%%

\bibitem{Liu}
H.~Liu,
\textit{``{Scattering in anti-de Sitter space and operator product
  expansion}''},
\doiref{10.1103/PhysRevD.60.106005}{Phys.~Rev. \textbf{D60}, 106005 (1999)},
\normalsize{\texttt{\arxivref{hep-th/9811152}{hep-th/9811152}}}.
%%CITATION = HEP-TH/9811152;%%

\bibitem{AdSfromCFT}
A.~L. Fitzpatrick \& J.~Kaplan,
\textit{``{AdS Field Theory from Conformal Field Theory}''},
\doiref{10.1007/JHEP02(2013)054}{JHEP \textbf{1302}, 054 (2013)},
\normalsize{\texttt{\arxivref{1208.0337}{arXiv:1208.0337}}}.
%%CITATION = ARXIV:1208.0337;%%

\bibitem{HartmanExcitedStates}
C.~T. Asplund, A.~Bernamonti, F.~Galli \& T.~Hartman,
\textit{``{Holographic Entanglement Entropy from 2d CFT: Heavy States and Local
  Quenches}''},
\normalsize{\texttt{\arxivref{1410.1392}{arXiv:1410.1392}}}.
%%CITATION = ARXIV:1410.1392;%%

\bibitem{KeskiVakkuri:1998nw}
E.~Keski-Vakkuri,
\textit{``{Bulk and boundary dynamics in BTZ black holes}''},
\doiref{10.1103/PhysRevD.59.104001}{Phys.~Rev. \textbf{D59}, 104001 (1999)},
\normalsize{\texttt{\arxivref{hep-th/9808037}{hep-th/9808037}}}.
%%CITATION = HEP-TH/9808037;%%

\bibitem{Dong:2014tsa}
X.~Dong, D.~Z. Freedman \& Y.~Zhao,
\textit{``{Explicitly Broken Supersymmetry with Exactly Massless Moduli}''},
\normalsize{\texttt{\arxivref{1410.2257}{arXiv:1410.2257}}}.
%%CITATION = ARXIV:1410.2257;%%

\bibitem{Kraus:2002iv}
P.~Kraus, H.~Ooguri \& S.~Shenker,
\textit{``{Inside the horizon with AdS / CFT}''},
\doiref{10.1103/PhysRevD.67.124022}{Phys.~Rev. \textbf{D67}, 124022 (2003)},
\normalsize{\texttt{\arxivref{hep-th/0212277}{hep-th/0212277}}}.
%%CITATION = HEP-TH/0212277;%%

\end{thebibliography}

\end{document}